\documentclass{aa}

\usepackage{graphicx}
\usepackage{txfonts}
\newcommand{\cd}{d$^{-1}$}

\begin{document}

\title{Oscillating Blue Stragglers, $\gamma$ Doradus stars and eclipsing
binaries in the open cluster NGC2506
\thanks{Based on observations obtained at the Flemish Mercator telescope on
La Palma, Spain, the Danish 1.5-m telescope at ESO, La Silla, Chile, and
on observations collected at the European Southern 
Observatory, Paranal, Chile (ESO Programme 075.D-0206(B))}}

\author{T. Arentoft\inst{1,2}
   \and J. De Ridder\inst{3}
   \and F. Grundahl\inst{1,2}
   \and L. Glowienka\inst{1}
   \and C. Waelkens\inst{3}
   \and M.-A. Dupret\inst{4}
   \and A. Grigahc{\`e}ne\inst{5} \and \\
   K. Lefever\inst{3}
   \and H.R. Jensen\inst{1}
   \and M. Reyniers\inst{3}
   \and S. Frandsen\inst{1,2}
   \and H. Kjeldsen\inst{1,2}
}

\offprints{T. Arentoft}

\institute{Department of Physics and Astronomy, University of Aarhus, 8000 Aarhus C., Denmark
           \email{toar@phys.au.dk}
\and Danish AsteroSeismology Centre, University of Aarhus, 8000 Aarhus C., Denmark 
      \and Instituut voor Sterrenkunde, Katholieke Universiteit Leuven, 3001 Heverlee, Belgium
\and LESIA, Observatoire de Paris-Meudon, UMR 8109, 92190 Meudon, France
\and GRAAG, Algiers Observatory, BP 63 Bouzareah, 16340 Algiers, Algeria 
}

\date{Received 14 December 2006 / Accepted 25 January 2007}

\abstract
{This is the first step in a project to combine studies of 
eclipsing binaries and oscillating stars to probe the interior of 
Blue Stragglers (BS). This may imply a way to discriminate 
observationally between different birth mechanisms of BS stars.
}
{We study the open cluster NGC2506 which contains oscillating BS 
stars and detached eclipsing binaries for which accurate parameters 
can be derived. This will tightly constrain the cluster isochrone 
and provide an absolute mass, radius and luminosity-scale for the cluster 
stars along with the cluster age, metallicity and distance. The present work
focuses on obtaining the light curves of the binaries and
determine their orbital periods, on obtaining power spectra of 
the oscillating BS stars to select targets for follow-up studies, and on 
searching for $\gamma$~Doradus type variables which are also expected to 
be present in the cluster. 
}
{With a two-colour, dual-site photometric campaign we obtained 3120 CCD-images 
of NGC2506 spread over four months.
We analysed the $BI$ time-series of the oscillating
stars and used simulations to derive statistical uncertainties of the 
resulting frequencies, amplitudes and phases. A preliminary mode-identification 
was performed using frequency ratios for the oscillating BS stars, and 
amplitude ratios 
and phase differences for a population of newly detected $\gamma$ Doradus stars.
}
{We quadrupled the number of known variables in NGC2506 by discovering 3 new
oscillating BS stars, 15 $\gamma$ Doradus stars and four new eclipsing 
binaries. The orbital periods of 2 known, detached eclipsing binaries were 
derived. We discovered a BS star with both p-mode and g-mode variability
and we confronted our $\gamma$~Doradus observations with state-of-the-art 
seismic 
models, but found significant discrepancy between theory and observations.
}
{
NGC2506 is an excellent target for asteroseismic tests of stellar models,
as strong external constraints can be imposed on the models of a population 
of more than 20 oscillating stars of different types. 
}
\keywords{stars: blue Stragglers, stars: variables: delta Sct,
stars: binaries: eclipsing - open clusters and associations: individual: 
NGC2506 - techniques: photometric}

\titlerunning{Oscillating Blue Stragglers, $\gamma$~Dor stars and EBs in NGC2506}

\maketitle

\section{Introduction}

Blue straggler (BS) stars are still a much debated phenomenon. In a 
cluster HR diagram, these stars are situated above and blueward of 
the main sequence (MS) turn-off point, where more massive stars 
were located before they evolved away to red giants. Their
appearance in a region of the HR diagram where they should not be
contradicts our standard single-star evolution theory.

Although the origin of BS stars is still not fully understood, 
the generally accepted working hypothesis is that these stars 
originate from stellar collisions, binary
coalescence, or from mass transfer in a binary system. These ideas go back
to the pioneering work of Hills \& Day (1976), and McCrea (1964). 
Because of these stellar interactions, a BS star is formed with a mass 
greater than the turn-off mass of the cluster. Such a star is therefore 
able to prolong its life on the MS, or even reappear as a born-again star, 
which explains why BS stars have not yet evolved far away from the MS.

However, before this hypothesis can be promoted to a theory, more 
observational and theoretical evidence needs to support and quantify it, 
and this is what current BS research is focusing on. We mention the recent 
result of Ferraro et al.~(2004) who found a bimodal radial BS population 
distribution in the globular cluster 47 Tuc, which they interpreted as 
evidence for different but simultaneously working BS birth mechanisms 
(see also Mapelli et al. 2006 for a follow-up article).
For the same cluster Ferraro et al.~(2006) pointed out that at least some 
of the BS stars seem to show the chemical signature of binary mass 
transfer, adding another piece of evidence for the mass transfer scenario. 
Progress on understanding the collisional scenario has mainly been made on 
the theoretical side. Sills et al.~(2001, 2005) used
smoothed particle hydrodynamics simulations to show that off-axis collision 
products have rotational velocities well above their break-up velocity so 
that they can only survive if they can somehow get rid of the excess angular 
momentum. The authors mention magnetic breaking caused by a circumstellar 
disk as a possible cause of a BS spin down. Such discs may 
(De Marco et al. 2004) or may not (Porter \& Townsend 2005) already have 
been discovered around BS stars.

Clearly, the theories for the different BS birth mechanisms are getting 
more established 
through the increasing amount of evidence. It's still difficult, though, 
to recognize how a particular BS star was born. Especially a diagnostic to 
discriminate observationally between binary coalescence and stellar 
collision is not available at the moment. This makes it difficult to study, 
for example, to what extent and for how long the interior of a BS star 
formed by ``unrolling'' a donor star is different from one formed by a 
direct collision of two stars.

Might asteroseismology help in this respect? In many clusters the BS stars 
are situated in the classical instability strip where one expects $\delta$ 
Scuti or SX Phe oscillations. Many oscillating BS stars have indeed already 
been detected, see for example~Gilliland \& Brown (1992), 
Gilliland et al.~(1998).
These oscillations could provide a way to model the BS star interior. The 
fact that the BS stars are part of a cluster makes them even more suitable 
asteroseismic targets, as chemical composition and age may be assumed 
constant for each BS member. These constraints can be made quantitatively 
more stringent if the cluster also contains detached eclipsing binaries for 
which accurate masses and radii can be determined, so that the isochrone can 
be precisely calibrated (see Grundahl et al. 2006). 
Finally, a combined set of observed frequencies in a number of oscillating 
stars within the same cluster, can be used to extract more information than 
for single stars (Frandsen \& Kjeldsen 1993).

This paper reports on the first stage of this Blue Straggler project: the 
challenging task of deriving useful oscillation spectra of BS 
stars in the open cluster NGC2506.
The reason for choosing NGC2506 is fourfold. First, the cluster contains
BS stars of which three of them were shown to oscillate by Kim et al.~(2001), 
by means of photometric time series (V1--V3). Secondly, the cluster contains 
detached eclipsing binaries. 
One was found by Kim et al. (2001; V4), and another by ourselves (V5).
Thirdly, it is an open cluster close to the celestial equator so that it is 
visible from both hemispheres, making a multi-site campaign of variable stars
possible. Finally, most of the cluster is confined within a radius of 6 arcmin,
making it fit nicely in a typical CCD image; yet the cluster is not 
so crowded that the photometric precision is affected. 

Besides the BS project, we present another project which started as a 
side result but which has meanwhile grown into an equally important project. 
The turn-off stars of NGC 2506 span a large 
part of the $\gamma$~Doradus instability strip making the cluster perfect for 
investigating the $\gamma$~Doradus phenomenon, should such stars be present.  
Kim et al.~(2001) searched for $\gamma$~Doradus stars without detecting any.
The data presented here are both more extensive and have a higher
precision per data point, making the search more sensitive and, as it turns
out, more successful because we do detect a surprisingly high number 
of $\gamma$~Doradus 
stars in the cluster. These $\gamma$ Doradus stars are very promising 
asteroseismic objects as their high-order g-modes probe deeply in the stellar 
interior. In addition, they are exciting laboratories to study stellar
physics such as semi-convection, convective overshooting, and (both slow 
and fast) stellar rotation.

Only recently, however, could some quantitative comparison between 
theory and observations be carried out for the $\gamma$~Doradus stars. 
Dupret et al.~(2004, 2005a) 
explained the driving mechanism of the $\gamma$~Doradus g-modes by using the
time-dependent convection (TDC) treatment of Grigahc\'ene et al. (2005).
With the same TDC models, Dupret et al.~(2005b) explained
for the first time and with success the 
photometric amplitude ratios and phase differences of five field $\gamma$ 
Doradus stars. In the same way as for the BS stars, our cluster $\gamma$ Doradus
stars provide a more stringent test of the theory. This paper 
reports the discovery of many new $\gamma$ Doradus stars, the observed 
frequencies, and a first attempt to model the photometric amplitude 
ratios and phase differences with the TDC theory, with remarkable results.

\section{The target cluster: NGC2506}

\label{targetcluster}
NGC2506 ($\alpha_{\rm 2000},\delta_{\rm 2000} = 08^{\rm h}00^{\rm m}01^{\rm s}$, $-10^{\degr}46^{\arcmin}12^{\arcsec}$)
is an old open cluster that already received quite some attention in the 
literature. 
The morphology study of Chen et al.~(2004) indicates that it is a mildly 
elongated cluster counting about 1091 stars, and the recently published 
catalogue of Kharchenko et al.~(2005) puts NGC2506 at a distance of 3460 pc. 
There seems to be a consensus in the literature
about the age of the cluster, which would be $t = 2.14 \pm 0.35$ Gyr 
(Salaris et al. 2004) although Xin \& Deng (2005) quotes an age of 3.4~Gyr.
Such an old age is inconsistent with our Str{\"o}mgren photometry
which also points to an age close to 2~Gyr.
We refer to Carretta et al.~(2004) for a literature overview on the 
metallicity of NGC2506. Although the different values vary quite a lot, 
all chemical analyses show that the cluster is metal deficient. We give a 
somewhat higher weight to the value obtained by Carretta et al.,
[Fe/H] = $-0.20\pm0.01$, as it is the only metallicity obtained through 
an iron abundance analysis using high-resolution spectroscopy. Nevertheless,  
we treat their quoted error 
bar as a lower limit, as their average metallicity value is based on a 
sample of only 2 stars. Also, their value differs from
other values reported in the literature obtained from photometry or 
low-resolution spectroscopy. 
Gratton (2000), for example, lists a value of [Fe/H] = $-0.42\pm0.09$. 
This leads us to believe that the uncertainty on the
metallicity is actually larger than 0.1 dex. As a final cluster parameter, 
we adopt the reddening given by Carretta et al.~(2004): 
E(B-V) = $0.073\pm0.009$.

\section{The data and the data reduction}

Our photometric time-series CCD observations were obtained
using DFOSC at the Danish 1.54-m telescope at ESO, La Silla, 
and MEROPE at the Flemish 1.2-m Mercator 
telescope at La Palma. At ESO we observed during 23 nights in January 2005, 
at La Palma during 20 nights in the period January--April 2005. 
Overlapping data were obtained on some of the January nights. 

In order to optimise the time series to the highest possible precision, 
care was taken to maintain the same pointing during the observations, 
i.e., to keep the stars at fixed pixel positions on the CCD. This was 
done by starting each night with the same pointing as the previous nights
(within a few pixels) and keeping this pointing using autoguiding. 
The exposure times
were chosen so that the brightest of the known variables, V1, 
were below the saturation limit. At ESO, this led to exposure times of 
approximately 100 sec in $B$ and 30 sec in $I$, hence sequences of 
(2$B$, 4$I$, 2$B$, ...) were obtained. At the Mercator, longer exposure 
times were possible due to the smaller telescope aperture, but 120 sec were 
used for both filters in order to maintain sufficient time resolution 
for the short-period oscillating BS stars.

\begin{figure} 
\resizebox{\hsize}{!}{\includegraphics{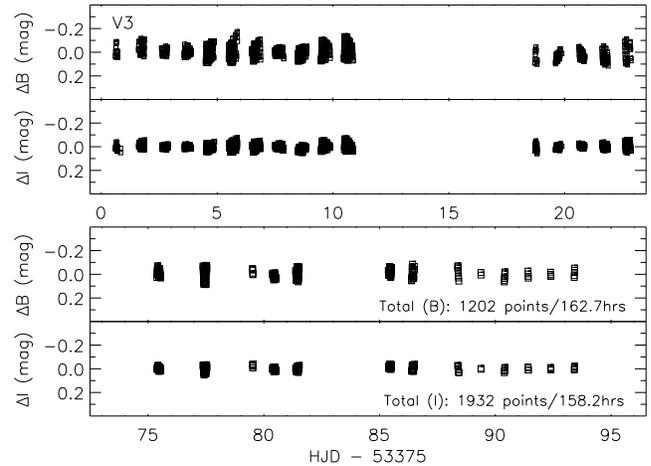}}
\caption{The total data set for the variable V3, showing the time 
distribution of the combined data from the two sites (ESO and La Palma).
\label{fig.V3total}}
\end{figure}

In total, 2100 $BI$ images were obtained at ESO, and 1020 at La Palma,
corresponding to 101.8 ($B$) and 97.8 ($I$) hours of observing time
at ESO, and 60.9 ($B$) and 60.4 ($I$) at La Palma.
In Fig.~\ref{fig.V3total} all data are shown for the
variable V3, illustrating the time distribution of the data set. 
Most of the data were obtained in January 2005 (136/132 hours 
for $B$/$I$) with additional data obtained in March and April at the 
Mercator. In the gap between the first and second batch of observations 
(Fig.~\ref{fig.V3total}, upper panels) a few data points were obtained
per night at La Silla, to monitor the EBs. These data are not included in the
analysis of the oscillating stars presented here. The data from March and
April were obtained at higher airmass and had lower exposure levels 
(\#counts/star), resulting in higher noise per data point.

The CCD images were calibrated using standard procedures.
The mean BIAS level was subtracted from each frame using overscan areas, 
and a median BIAS image (with zero mean) was then subtracted.
Evening and morning sky-flats had been obtained whenever possible, and from
those, a single, median sky-flat were constructed per filter and per observing 
run. The run at the D1.5-m spans nearly a month, but before applying the 
flat-field calibration to the science images,
we checked that the flat-fields were indeed stable over the length of 
the observing run. 

The photometric reductions were done separately for the Mercator and ESO 
data, using the software package MOMF (Kjeldsen \& Frandsen 1992).
MOMF applies a very robust algorithm combining PSF and aperture photometry,
and is excellent for time-series photometry in semi-crowded fields. 
The data from each site were merged on a star-by-star basis, by shifting 
the mean levels of the individual light curves to zero.  
An example of this merging is shown in Fig.~\ref{fig.V3overlap} for V3 during 
a single night. Not all stars were present in both data sets; due to 
differences in field-of-view (MEROPEs FOV is 
6$\stackrel{'}{\textstyle .}$5$\times$6$\stackrel{'}{\textstyle .}$5, 
DFOSCs 15$'\times$15$'$) more stars were observed at ESO. The observed 
field is shown in Fig.~\ref{fig.finding} (a DFOSC image), which also serves 
as a finding chart for the variable stars discussed below.

\subsection{Analysis of differential light curves}
 
The output from MOMF is differential time series photometry for all stars 
in the images, using the mean of all stars as the reference
level to take out atmospheric effects. The light curves were kept in the 
instrumental system, but calibrated 
Str{\"o}mgren photometry has previously been obtained for the cluster by
F. Grundahl; these data are presently unpublished, but calibrated as 
described in Grundahl et al. (2002).
We will in the present paper rely on these data for deriving stellar
parameters for the variable stars.

We obtained light curves for 863 stars in the MEROPE field and
3354 stars in the DFOSC field, of which one must expect that a significant 
fraction, especially of those far from the cluster center, are not physical   
members of the cluster. The resulting photometric precision in the light 
curves was as aimed for, with a precision per data point of 1--2 mmag for 
the bright stars (on good nights), gradually decreasing for stars of fainter 
magnitudes as the photon noise increases.

\begin{figure} 
\resizebox{\hsize}{!}{\includegraphics{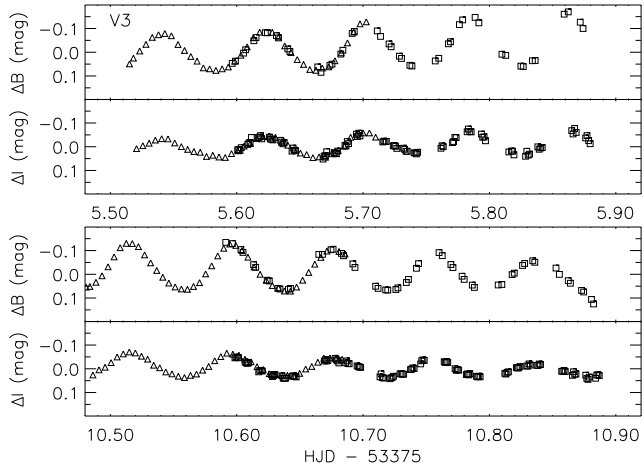}}
\caption{Overlapping data on V3.
Triangles are data from La Palma, squares are from 
La Silla. A slow trend with a frequency of 1.0~\cd~was present in the 
La Silla $B$-data only, and was subtracted before merging the data 
(see Sections 3.1 and 4.1).
\label{fig.V3overlap}}
\end{figure}

The aim of our data analysis was to detect all variable stars in the 
observed fields, and to use all data for a detailed analysis of the 
variability. The first step in the analysis was to clean all light curves for 
obvious outliers and to cross-identify the common stars in the MEROPE and 
DFOSC fields. We then calculated amplitude spectra for all stars, based 
on the MEROPE and DFOSC data separately and inspected the light curves and
amplitude spectra of each star in the two data sets manually, to select 
possible variable stars for further analysis. For stars to be selected as
candidates, we required that variability was present in both the MEROPE 
and DFOSC data (for common stars) and in both the $B$- and $I$ filter data.

At this stage we had a list of 33 variable star candidates, of which 28 
ended up being retained as variables, while 5 stars were rejected at a 
later stage in the analysis.
For the candidates observed at both telescopes we now merged the MEROPE and 
DFOSC data by shifting the light curves to zero mean. For some of the 
blue stragglers (such as V3; see Fig.~\ref{fig.V3overlap}), we subtracted 
some incoherent, low-frequency
variations in order to match up the data from the two sites. As these stars 
are bluer than the large majority of cluster stars, we ascribe these 
variations to second-order extinction effects that remain uncorrected
because the differential photometry is dominated by the redder cluster 
turn-off stars. 

Because the data quality during a time series is variable, we assigned 
statistical weights to each data point (see e.g., Arentoft et al. 1998, 
Handler 2003). This was 
done by highpass filtering the time series, using the Fourier transform, so 
that the filtered light curve only contained high-frequency (white) noise, 
which is useful for finding deviating points. We tried different weighting 
schemes based on the deviation of the individual data points from the light 
curve mean compared to the overall standard deviation ($\sigma$)
of the light curve. In the end we chose to give points deviating 
less than 2$\sigma$ in the highpass filtered series weight 1.0, while 
points deviating more were weighted as 1/$\sigma^2$, as this
scheme resulted in the lowest noise in the weighted amplitude spectra.  

\begin{figure} 
\resizebox{\hsize}{!}{\includegraphics{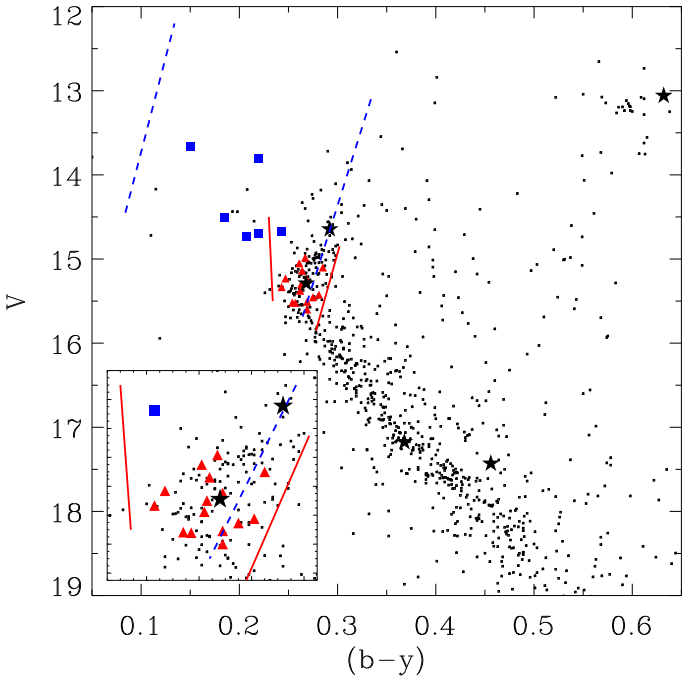}}
\caption{Colour-magnitude diagram for NGC2506 with the detected variables
marked. The blue dashed lines are the observational $\delta$~Scuti 
instability strip (Breger 2000) and the red solid lines the $\gamma$~Doradus 
instability strip from Handler \& Shobbrook (2002). A distance modulus of 
12.65 and $E(b-y)=0.054$ was used for placing the instability 
strips in the CMD. Blue squares marks the oscillating BS stars, red triangles the 
$\gamma$~Doradus candidates, and black 
stars the eclipsing binaries. The insert in the lower left corner shows a 
zoomed view of the $\gamma$~Dor instability strip.
(Colours are only visible in the electronic version). 
\label{fig.cmd}}
\end{figure}

We then scrutinized the evidence for variability, using weighted amplitude 
spectra as well as the light curves, for each of our 33 candidates. In this
way we detected 28 variables, of which only 5 were previously known. The 
variables, which will be discussed in detail below, are 6 oscillating blue 
straggler stars ($\delta$~Scuti oscillations), 15 $\gamma$~Doradus candidates,
6 binaries and one star of presently unknown type. The classification is 
based on the time scale of the variability ($\sim$days for the $\gamma$~Doradus
candidates, $\sim$hours for the oscillating blue stragglers), $BI$ amplitude 
ratios -- to separate $\gamma$~Doradus stars from, e.g., ellipsoidal 
variables -- as well as the 
position of the stars in the colour-magnitude diagram. Fig.~\ref{fig.cmd} plots
this diagram based on the Str{\"o}mgren data mentioned above, with the 
detected variables and the observational $\delta$~Scuti (Breger 2000) and 
$\gamma$~Doradus (Handler \& Shobbrook 2002) instability strips indicated. 
In fact, we will 
in the following argue, that the majority of the $\gamma$~Doradus candidates,
which are all positioned inside the instability strip, 
can be considered {\it bona fide} $\gamma$~Doradus stars (Handler 1999, 
Handler \& Shobbrook 2002).

The last step in the data analysis was a detailed frequency analysis of the
oscillating stars, and a period search for the eclipsing binaries. For the 
oscillating stars, the frequency analysis was carried out using {\sc Period98}
(Sperl 1998), which uses simultaneous least-squares fits of all detected 
frequencies in a light curve.  
We only include frequencies that are present in both the $B$ and the $I$ data, 
with a Signal-to-Noise ratio in the amplitude spectra of 4 in at least one 
of the filters, although most detected
frequencies meet this requirement for both filters. Because our $B$ and $I$ 
data have different sampling, the spectral window functions are slightly
different for the two cases. 
This may lead to small differences in the detected frequency values between 
the two filters. For each detected frequency, we therefore manually inspected 
the amplitude spectra, to find a frequency value that described the 
variation in both filters well. For each star, we then kept the detected
frequencies fixed, and fitted all frequencies simultanously to the $B$ and
$I$-band data separately, to determine amplitudes and phases for both 
filters. 

All detected frequencies for the oscillating blue stragglers and the
$\gamma$~Doradus stars are given in Tables~\ref{tab.bs} and~\ref{tab.gd},
and Str{\"o}mgren indices for all variables, as well as stellar parameters 
for most of these stars, have been collected in Table~\ref{tab.param}. We have 
used the calibration of Alonso et al. (1996) for deriving effective 
temperatures, the bolometric corrections follow 
Lejeune et al. (1998), and luminosities have been derived using the known 
cluster distance and $M_{bol,\odot}=4.746$.

\subsection{Mode parameter uncertainties}

As we plan to use the amplitude ratios and phase differences for theoretical 
modeling of the $\gamma$~Doradus stars, we should have an idea how large 
the uncertainty is on the observed values. One possibility would be to use 
the well-known expressions for the standard errors of the amplitudes
and phases given by e.g.~Breger et al.~(1999). One should be aware, however, 
that these expressions were derived for equidistant time series, and under 
the assumption that the true oscillation frequencies are known. Our time 
series are notoriously non-equidistant. In addition, as many of our targets 
are multi-periodic, we can expect that the derived frequencies for the 
lowest-amplitude modes may deviate somewhat from the true eigenfrequencies.
Hence, the standard formulas for calculating errors are not directly 
applicable to our data set.

We therefore used extensive simulations to obtain uncertainty estimates. For 
each star we simulated $B$ and $I$ lightcurves with exactly the same time 
sampling as for the observed time series. Each time series consisted of a 
sum of sines with exactly the same frequencies, amplitudes and phases as 
derived from the observed time series. Next, white noise was added
with the same rms value as in the residuals of the observed time series. 
These time series were then subjected to an automated analysis procedure 
that mimicked accurately how we analysed the observed time series. 
Oscillation frequencies were searched for in the $B$ time series with 
cyclic prewhitening, and these frequency values were then used to perform 
a least-squares fit of the $B$ and $I$ time series to obtain the amplitudes 
and phases.  The frequency analysis could be automated 
because the algorithm searched in predefined intervals 
$[\nu_i - \frac{2}{T}, \nu_i + \frac{2}{T}]$ where
$\nu_i$ is the $i$-th observed frequency (used as input to construct 
the synthetic time series) and T is the total time span of the time series. 
From the result of this fit, amplitude ratios and phase difference could be 
computed. For each star, the procedure was then repeated
4000 times. The result is for each oscillation mode of each star, a set 
of 4000 amplitude ratios and a set of 4000 phase differences, one for each 
noise realisation in the time domain. Note that because of the setup with 
the predefined intervals, the automated procedure cannot
be fooled by one-day aliases. It can still be misled, however, by alias 
frequencies smaller than $2/T$. 

There are two questions we would like to answer with these simulations. 
First, can we retrieve the frequencies, amplitudes and phase that were 
used as input to construct the synthetic time series? This is a necessary 
condition to have confidence in the observational results, which are listed 
in Tables~\ref{tab.bs} and~\ref{tab.gd}. The short answer is ``yes''. The 
sets of simulations can be used to construct histograms of the fit 
parameters, which in turn can be used to compute average values and standard 
deviations. The a posteriori derived mean frequencies are always very 
close to the input frequencies, which is reassuring. Sometimes, however, 
the histogram is bimodal or even trimodal with well separated peaks. For 
these frequencies the problem of aliasing is very severe, and this should 
be taken into account when determining the uncertainties. 

\begin{figure}
\resizebox{\hsize}{!}{\includegraphics{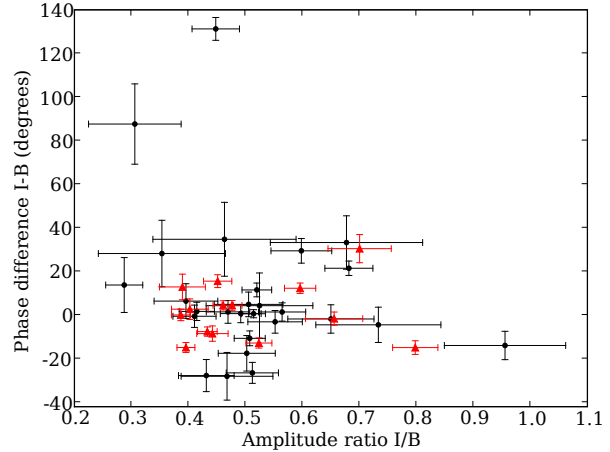}}
\caption{The amplitude ratio $I/B$ versus the phase difference $I-B$ for all 
oscillation frequencies detected in the $\gamma$~Doradus stars. The 
coordinates of the bullets are the observed values, the error bars
were computed from the simulated time series. Red triangles are used for those 
highest-amplitude modes for which no aliasing is observed in the simulated time series.\label{ratiodiff}}
\end{figure}

Second question: how large are the uncertainties of the observed amplitude 
ratios and phase differences? We use the variances derived from the simulated 
datasets as an uncertainty measure for the observed quantities. Always 
the total variance, i.e. intra-variance + inter-variance, was computed 
taking into account possible confusion with window aliases. The results are 
shown in Fig.~\ref{ratiodiff} for the $\gamma$~Doradus stars. For every 
target we checked whether the highest-amplitude mode
showed aliasing in the simulated data (i.e.~a multi-modal histogram) and 
designated the ones that don't, with red triangles in the figure. We regard 
the observational amplitude ratios and phase differences of these latter 
modes as most reliable. As we expected, all our uncertainty estimates are 
comparable or somewhat larger than the
estimates we would have obtained with the formalism of Breger et al.~(1999).
We postpone futher discussion of Fig.~\ref{ratiodiff} to Section~5. 

As a final comment we should mention that for 3 modes ($\nu_1$ of V12, 
$\nu_2$ of V15, and $\nu_4$ of V17) the phase differences showed a weak 
trend with respect to the frequencies in the dataset of 4000 realisations.
This signals that we should treat the observational values for these 3 
modes with caution. 

\section{Oscillating blue straggler stars}

NGC2506 has a known population of 12 BS stars, which are all probable cluster
members based on proper motions (Xin \& Deng 2005, Chiu \& van Altena 1981). 
Kim et al. (2001) found 3 oscillating BS stars (V1--V3), of which one (V3), was 
not included in the proper motion study of Chiu \& van Altena. We discover 
an additional three oscillating BS stars (V6--V8) of which V7 has a 
membership probability of 42\% from the proper motion study, while the 
proper motions of the other two were not measured.
Although we do not have spectroscopic data to pinpoint their 
membership status, all six stars have Str{\"o}mgren indices in agreement 
with cluster membership -- their $m_1$ indices, for instance, are similar 
to those of the cluster turn-off stars. The positions of the six stars in 
the CMD are marked with blue squares in Fig.~\ref{fig.cmd} and the basic 
parameters are given in Table~\ref{tab.param}. Furthermore, the six stars have 
short oscillation periods (Table~\ref{tab.bs}, frequencies around 10--20~\cd), 
are mostly multiperiodic and, except V2, have low amplitudes (mmag). These are 
the signatures of $\delta$~Scuti type oscillations. Because the $\delta$~Scuti 
instability strip overlaps the region of the blue stragglers 
(Fig.~\ref{fig.cmd}) 
the six stars are most likely cluster members, as an instrinsically fainter 
foreground star, or a brighter background star, would not be expected to 
display $\delta$~Scuti type oscillations.  

\begin{figure} 
\resizebox{\hsize}{!}{\includegraphics{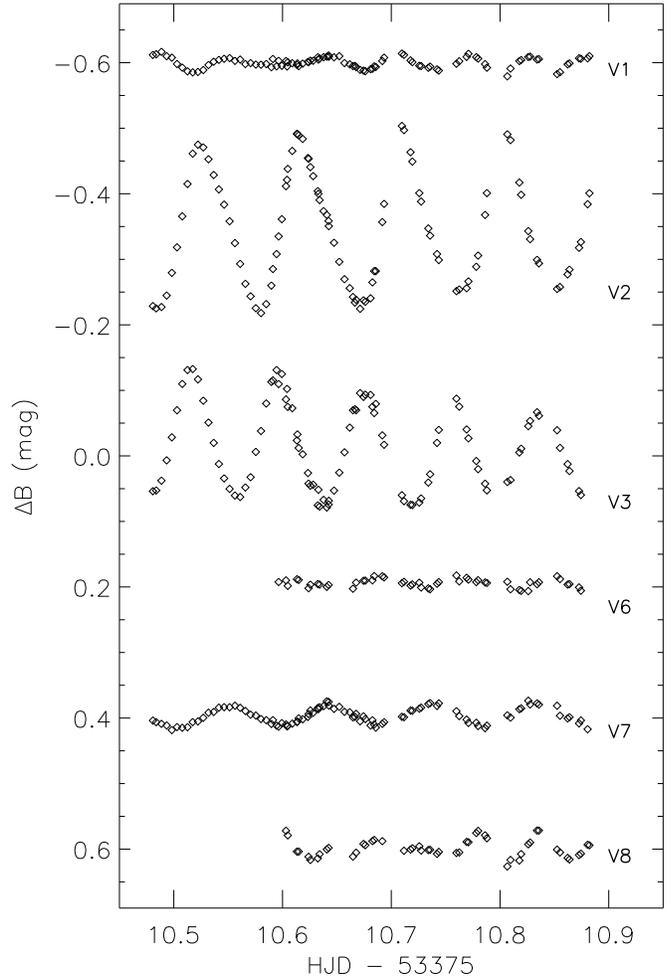}}
\caption{A single night of data on the BS stars. V6 and V8 were observed
from La Silla only
\label{fig.bslc}}
\end{figure}

Light curves from a single night for all 6 stars are shown in 
Fig.~\ref{fig.bslc}. V6 and V8 are outside the MEROPE FOV so we have only 
single site data for these stars. This results in a different
spectral window function, however for both the single- and the dual-site data, 
prominent sidelobes are present in the amplitude spectra. We show 
such spectra of V1--V3 in Fig.~\ref{fig.BSampl1} and
of V6--V8 in Fig.~\ref{fig.BSampl2}. The window function for the stars 
observed at both sites is best illustrated by V2, which has a single dominant 
frequency, while that of the single-site data is illustrated
by V6. For this star, the signature of the dominant frequency is, however, 
distorted by a low-amplitude mode at higher frequency. Below, we 
discuss the individual stars in more detail.

\subsection{The known variables V1--V3}

Kim et al. (2001) detected short-period oscillations in V1--V3, with 
dominant frequencies of 14.742, 10.854 and 12.264~\cd, respectively, in good
agreement with our results. They classified all three variables as 
$\delta$~Scuti stars, but only V1 as a blue straggler. However, based on
our new, more precise CMD (Fig.~\ref{fig.cmd}), the six short-period variables
are bluer than the cluster turn-off and should therefore all be classified as 
blue stragglers. Using a period-luminosity relation for $\delta$~Scuti stars,
Kim et al. arrived at a tentative identification for the dominant modes of V1 
and V2 as 3rd radial overtone and the fundamental radial mode, respectively. 

\begin{figure} 
\resizebox{\hsize}{!}{\includegraphics{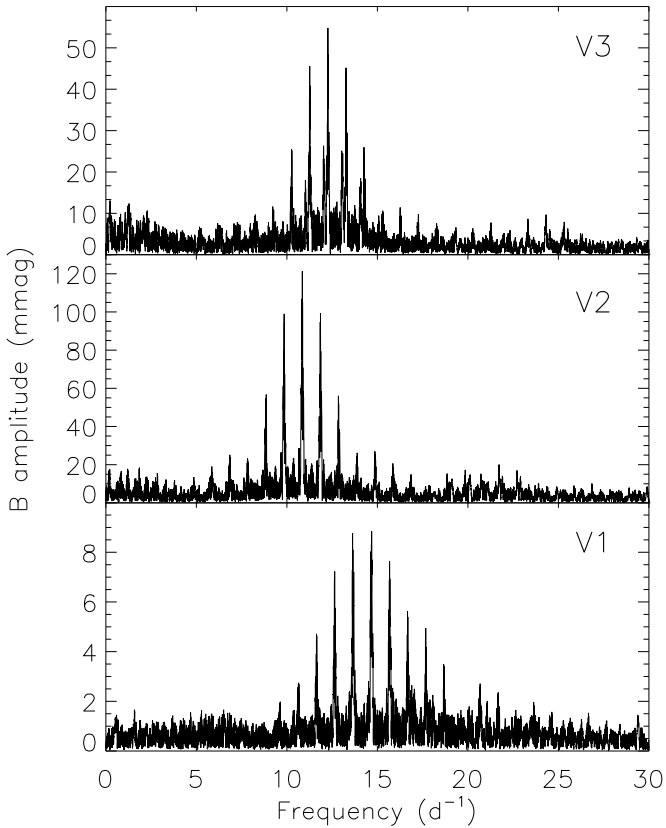}}
\caption{$B$ amplitude spectra of the three known, oscillating BS stars 
(V1--V3). Notice the differences in $y$-scale between the panels.
\label{fig.BSampl1}}
\end{figure}

V1 is the brightest and bluest star of the variables. The light curves of
this star were 
corrected for low-frequency (0--2~\cd) variations at the level of 
2--3~mmag before the data were merged. These variations caused minor 
mis-matches between the MEROPE and DFOSC data, and must therefore be 
ascribed to instrumental or atmospheric effects (see Sect.~3.1). 
In the final light curves, we detect 7 frequencies in the range 13--18~\cd, 
with amplitudes between 2 and 5 mmag (Table~\ref{tab.bs}). 
V2 is a high-amplitude variable (0.12~mag in $B$). The data from the two 
sites match up well, and no low-frequency filtering was necessary when merging
the data. 
The dominant mode at 10.854~\cd~is distorted as we detect 2$\nu_1$ and 
3$\nu_1$ as well. $\nu_2$ appears to be an independent term, while $\nu_3$ 
is equal to $\nu_1+\nu_2$. Such asymmetries and mode-couplings are common 
in high-amplitude $\delta$~Scuti stars. 

Some of the BS stars fall within the $\gamma$~Doradus
instability strip as well. V3 is located in the CMD just 
to the blue side of the $\gamma$~Doradus instability strip, but it is
nevertheless the only one of the six oscillating BS stars which
seems to display both low- and high-frequency variability. The fast variations
in V3 (Fig.~\ref{fig.V3overlap},~\ref{fig.bslc} and~\ref{fig.BSampl1} ) are 
clearly visible. We detect 8 frequencies (11--25~\cd), with 
amplitudes in the range 3--56~mmag in $B$. When merging the data from the 
two sites, it was necessary first to remove a relatively strong 
(25~mmag) signal at 1~\cd~from the DFOSC $B$-data only, as mentioned in 
the caption of Fig.~\ref{fig.V3overlap}. The other  
three data sets (MEROPE $BI$, DFOSC $I$) remained uncorrected. During the 
frequency analysis of the merged data, we found that three low-frequency 
terms were present in both the $B$- and $I$-data. The $I/B$ amplitude ratios 
of these frequencies are about 0.4, which is  very similar
to amplitude ratios of the $\gamma$~Doradus stars to be discussed below. 
This points toward oscillations as a possible explanation of the low-frequency 
variability in V3. This result should be confirmed, but it is an interesting
possibility that V3 may oscillate simultaneously in g-modes as the 
$\gamma$~Doradus stars, and in p-modes as the $\delta$~Scuti stars.  

\subsection{The new variables V6--V8}

\begin{figure} 
\resizebox{\hsize}{!}{\includegraphics{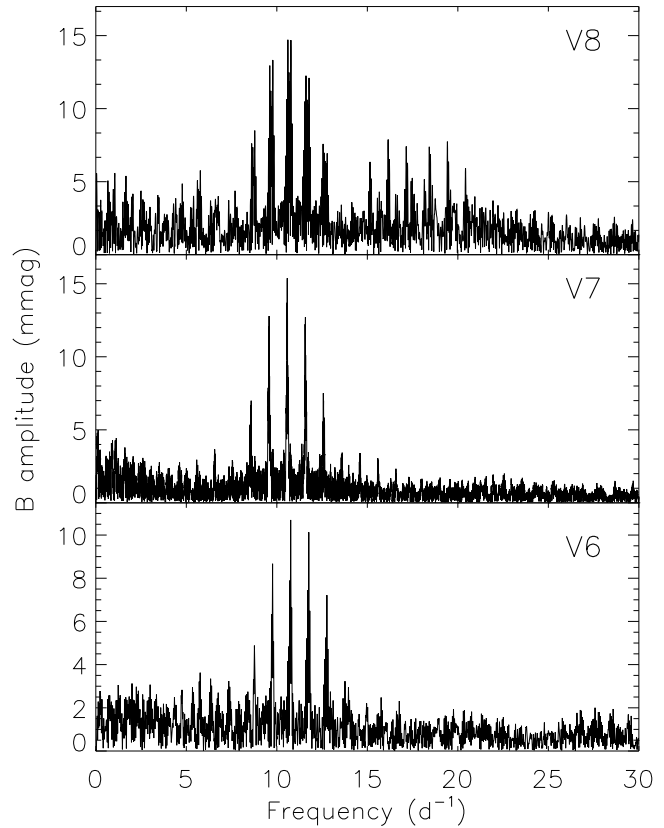}}
\caption{Amplitude spectra of the new oscillating BS stars V6--V8.
\label{fig.BSampl2}}
\end{figure}

In addition to the three oscillating BS stars discussed above, we
detected short-period oscillations in the three stars V6--V8 as well. The
amplitude spectra are shown in Fig.~\ref{fig.BSampl2}. \\
As mentioned above, we only have data from DFOSC on V6. As in the case of
V3, the $B$ data included a relatively strong signal at 1~\cd, which has 
been subtracted in Fig.~\ref{fig.BSampl2}. We detect three frequencies in
V6, although $\nu_1$ does not stand out clearly in the $I$ data. There 
appears to be further frequencies present in the range 10--15~\cd, but these
are below the 4$\sigma$~detection threshold of our data. Although V6 is 
located inside the $\gamma$~Doradus instability strip as well, we do not 
find evidence for low-frequency, $\gamma$~Doradus type variability.

For V7, we detect a single frequency, but the residuals after 
subtracting $\nu_1$ indicate that further frequencies may be
present -- especially a peak near 9.05 or 10.05~\cd~with a $B$-amplitude of
about 2~mmag just escapes a 4$\sigma$ detection in both the $B$ and $I$-data.
V8, on the other hand, is multiperiodic with 
at least 7 frequencies, although $\nu_7$ does not meet the 4$\sigma$ 
criterion for detection. It is, however, present in both the $B$- and 
$I$-filter data and is therefore included in Table~\ref{tab.bs}. Further 
frequencies are probably present in the light curves of V8, but escapes 
detection in our data. No low-frequency filtering has been applied to 
the V8-data shown in Fig.~\ref{fig.BSampl2}.

We finally mention the star V10, which displays variability
at the mmag-level with a periodicity near 10~\cd. This star is, however, 
positioned in the CMD near the giant stars of the cluster and is close
to the saturation limit of our data. Little can be said at this stage 
about the nature of this object, the variability seems $\delta$~Scuti like,
so it is probably not a cluster member. 

\begin{table}
\small
\caption{Variability data for the oscillating BS stars. 
Frequencies are in \cd~and amplitudes are given in mmag. The 
quoted errors on the frequencies are standard deviations on the 
values determined from extensive simulations (see text).
\label{tab.bs}}
\begin{tabular}{cccrrr}
\hline
\noalign{\smallskip}
ID & $\nu_{i}$ & $\nu$ & A$_{I}$ & A$_{B}$ & S/N$_{B}$ \\
\noalign{\smallskip} \hline
V1 & $\nu_{1}$ &13.630(1) & 3.2 & 6.8 &15.1    \\
   & $\nu_{2}$ &14.684(1) & 2.8 & 5.1 &11.2    \\ 
   & $\nu_{3}$ &14.716(1) & 1.7 & 2.9 & 6.4    \\
   & $\nu_{4}$ &16.688(5) & 1.5 & 2.8 & 6.2    \\
   & $\nu_{5}$ &16.896(2) & 1.0 & 2.2 & 5.0    \\
   & $\nu_{6}$ &17.043(2) & 0.9 & 1.8 & 4.0    \\
   & $\nu_{7}$ &17.672(1) & 2.2 & 4.5 & 10.0    \\
V2 & $\nu_{1}$ &10.854(1) &55.8 & 121.4&215.0  \\ 
   & $\nu_{2}$ &20.141(1) & 7.5 & 14.2 &29.5   \\
   &2$\nu_{1}$ &21.709(1) & 8.5 & 17.0 &35.3   \\
   & $\nu_{3}$ &30.995(2) & 1.2 &  2.8 & 5.9   \\
   &3$\nu_{1}$ &32.563(3) & 1.3 &  3.1 & 6.4   \\
V3 & $\nu_{1}$ & 0.526(1) & 3.2 &  6.9 & 3.1   \\
   & $\nu_{2}$ & 0.827(6) & 4.4 & 10.4 & 4.7   \\
   & $\nu_{3}$ & 0.908(3) & 4.9 & 12.2 & 5.5   \\
   & $\nu_{4}$ &11.243(2) & 1.7 &  5.8 & 6.4   \\
   & $\nu_{5}$ &12.031(1) & 7.4 & 16.8 &18.6   \\
   & $\nu_{6}$ &12.265(1) &25.9 & 56.2 &62.0   \\ 
   & $\nu_{7}$ &12.748(5) & 2.2 &  3.9 & 4.4   \\
   & $\nu_{8}$ &13.088(1) &10.0 & 18.2 &20.1   \\
   & $\nu_{9}$ &18.555(4) & 2.5 &  3.7 & 4.1   \\
   &$\nu_{10}$ &24.336(5) & 1.7 &  3.2 & 4.5   \\
   &$\nu_{11}$ &24.530(2) & 2.1 &  4.3 & 6.0   \\
V6 & $\nu_{1}$ & 9.441(3) & 0.9 &  3.7 & 5.9   \\
   & $\nu_{2}$ &10.767(1) & 5.1 & 10.9 &17.2   \\ 
   & $\nu_{3}$ &13.958(3) & 1.4 &  3.1 & 4.9   \\
V7 & $\nu_{1}$ &10.576(1) & 6.9 & 15.3 &22.3   \\ 
V8 & $\nu_{1}$ &10.227(2) & 2.9 &  4.5 & 4.5   \\
   & $\nu_{2}$ &10.611(1) & 4.4 & 11.1 &11.2   \\ 
   & $\nu_{3}$ &10.781(1) & 5.0 & 11.7 &11.7   \\
   & $\nu_{4}$ &11.550(2) & 4.7 &  8.3 & 8.4   \\
   & $\nu_{5}$ &16.149(2) & 3.4 &  6.8 & 6.8   \\
   & $\nu_{6}$ &19.423(2) & 3.4 &  7.2 & 7.2   \\
   & $\nu_{7}$ &20.944(3) & 1.8 &  3.4 & 3.4   \\
\hline
\noalign{\smallskip}
\end{tabular}
\end{table}

\subsection{Mode-identification of the oscillating BS stars}

\begin{figure} 
\resizebox{\hsize}{!}{\includegraphics{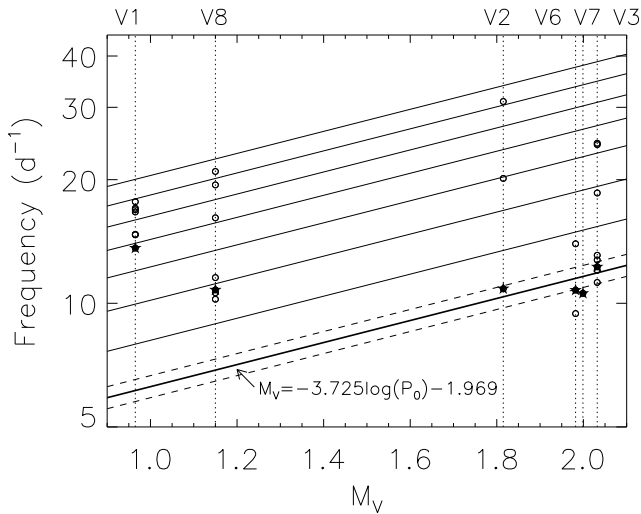}}
\caption{
Comparison between the observed frequencies in the BS stars and a 
period-luminosity relation for the fundamental mode of $\delta$~Scuti stars 
(thick solid line) and for consecutive radial overtones  
(solid lines) based on theoretical frequency ratios from stellar oscillation 
models. The dashed lines indicate the 1$\sigma$ error on the PLR and the 
vertical, dotted lines plots the $M_V$ of the individual variables. The 
star-symbols indicate the frequency of highest amplitude in each star.
\label{fig.bsplr}}
\end{figure}

Mode-identification in $\delta$~Scuti stars is a notorious problem. Unless
the modes can be identified directly, as in the case of the double-mode 
SX~Phe stars, or from combined spectroscopy and photometry for bright stars
such as FG~Vir (Viskum et al. 1998, Breger et al. 1999, Zima et al. 2006), 
one must turn to direct comparison with theoretical models. The latter method 
is further complicated 
in the present case, as the $\delta$~Scuti variables are blue stragglers, 
for which the formation history is unknown (see, e.g., Gilliland et al. 1998; 
Tian et al. 2006). In the 
case of NGC2506, cluster parameters and isochrones will be constrained from
the binary analysis of spectroscopic (VLT data are already obtained, and will
be published in a separate paper) and photometric (this paper) data 
on detached, eclipsing cluster binaries. In parallel with this, we will
initiate dedicated model calculations for the oscillating blue stragglers, 
but this is beyond the scope of the present paper. Instead, we will use a
period-luminosity relation (PLR) from Petersen \& Christensen-Dalsgaard (1999,
their Eq.~4), along with expected period ratios for consecutive, radial 
overtones in $\delta$~Scuti stars, to discuss the detected frequencies 
(see e.g., Petersen \& Christensen-Dalsgaard 1996, Gilliland et al. 1998). 
Although the period ratios should be evaluated using dedicated models, the 
ratios are quite constant between models of different mass (e.g., Fig.~21 in 
Gilliland et al. 1998). 
One should, however, be cautious when using period ratios for 
mode-identifications of low-amplitude, multiperiodic $\delta$~Scuti stars, 
as non-radial modes may easily show the same frequency ratios as the radial 
overtones, as noted by Poretti (2003). In addition, one should also be cautious
of the fact that for the present paper, we are using period ratios from
standard models of non-rotating $\delta$ Scuti stars for mode-identification
in BS stars. These assumptions should be kept in mind when we in the 
following discuss the mode-identification of the indvidual BS variables.
In Fig~\ref{fig.bsplr} we compare the observed frequencies with the 
period-luminosity relation for the fundamental (F) modes from 
Petersen \& Christensen-Dalsgaard (1999) and with consecutive radial overtones.

Kim et al. (2001) identified the dominant frequency of V1 as the third radial 
overtone (3O), while we find that this frequency ($\nu_1$) is better matched 
with the fourth radial overtone (4O). We note, by the way, that the frequency 
ratio between $\nu_1$ and $\nu_7$ is 0.771, which is the expected F/1O 
frequency ratio, 
but such an identification clearly disagrees with the PLR. Instead $\nu_7$ 
could possibly be 6O, but given the dense frequency spectrum, most, if not all,
frequencies are likely non-radial.  
The dominant frequency in the high-amplitude $\delta$~Scuti star 
V2 ($\nu_1$), on the other hand, is in good agreement with a F-mode 
classification, as also noted by Kim et al. The frequency $\nu_2$ could be 
identified as 3O 
if we allowed for a 1~\cd~shift to 21.141~\cd. Although we do have significant 
sidelobes in the amplitude spectra, neither the $B$ or $I$ data support such
a shift.   

The strongest signal in V3, $\nu_6$ (at 56~mmag amplitude in $B$), also
agrees well with the fundamental mode based on the PLR, while all the
other frequencies probably are non-radial. Several frequencies cluster near 
12~\cd, and three of them ($\nu_4,\nu_5,\nu_7$) form a near equidistant 
triplet. However, the error simulations puts the largest uncertainty on 
$\nu_7$, of 0.005~\cd, much smaller than the observed deviation from 
equidistance of 0.071~\cd. If V3 is a moderate to fast rotator, significant 
deviations from 
simple triplets are expected (Kjeldsen et al. 1998), but such conclusions 
would require comparison with dedicated stellar models as well as a 
spectroscopically derived rotational velocity.

V6 represents the best case in terms of mode-identification.
The frequency of the dominant mode ($\nu_2$) agrees with a F-mode 
classification, and the ratio between $\nu_2$ and $\nu_3$ is 0.771, which 
suggests a F/1O identification. $\nu_1$ must then be non-radial.
V7 is monoperiodic, but the single frequency is in agreement with the 
fundamental mode predicted by the PLR, within the uncertainties.
We do not have Str{\"o}mgren indices for V8, but $M_V \approx
1.15$ based on the $V$-magnitude. The dense spectrum near 10~\cd~clearly
suggests non-radial oscillations, and indeed, we have no evidence for 
radial modes.

To summarize, comparison with the PLR and expected frequency ratios 
suggest radial modes to be excited in V1 (4O), V2 (F), V3 (F), V6 (F/1O) 
and V7 (F).

\section{Discovery of $\gamma$~Doradus stars}

Among the new variable stars is a population of 15 stars displaying 
long-period variability on time scales from hours to 
days. The amplitudes of variability are in the range 5--30 mmag
and several stars are multiperiodic. They are positioned inside the 
$\gamma$~Doradus instability strip in the CMD (Fig.~\ref{fig.cmd}) which, 
along with the periods and amplitudes 
suggest that they may be $\gamma$~Doradus stars (Handler 1999; 
Handler \& Shobbrook 2002; Henry et al. 2005). We will in the 
following argue that these stars, labeled V11--V25, can in fact be 
classified as $\gamma$~Doradus stars based on the evidence presented below.

\subsection{Evidence for multiperiodic variability}

\begin{figure} 
\resizebox{\hsize}{!}{\includegraphics{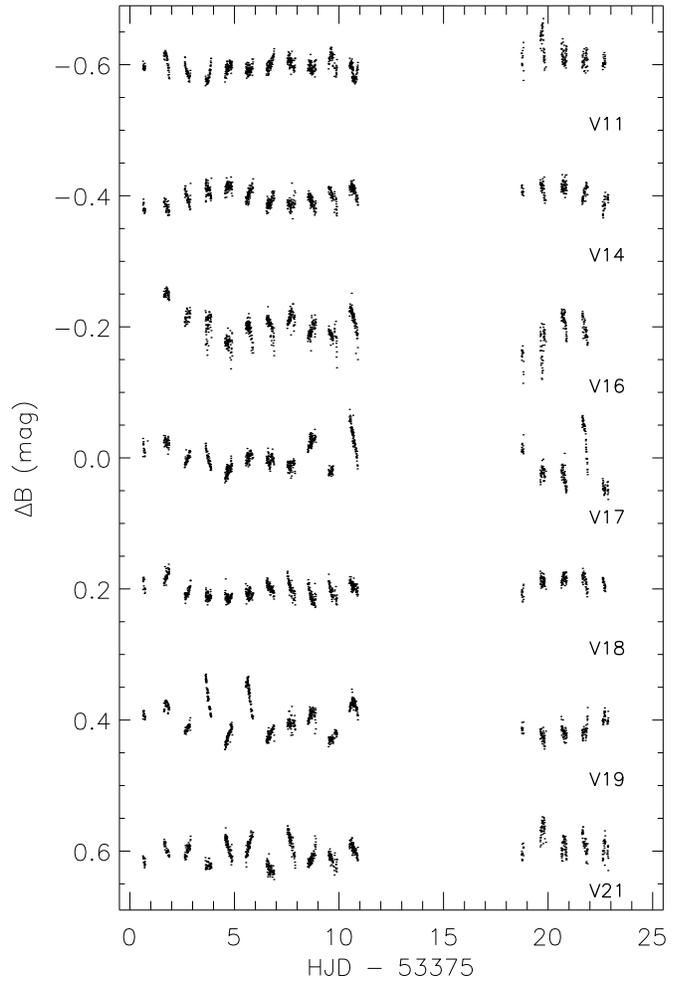}}
\caption{Light curves for 7 of the long-period variables ($\gamma$~Doradus)
stars in our sample.
\label{fig.gdlc}}
\end{figure}

\begin{figure} 
\resizebox{\hsize}{!}{\includegraphics{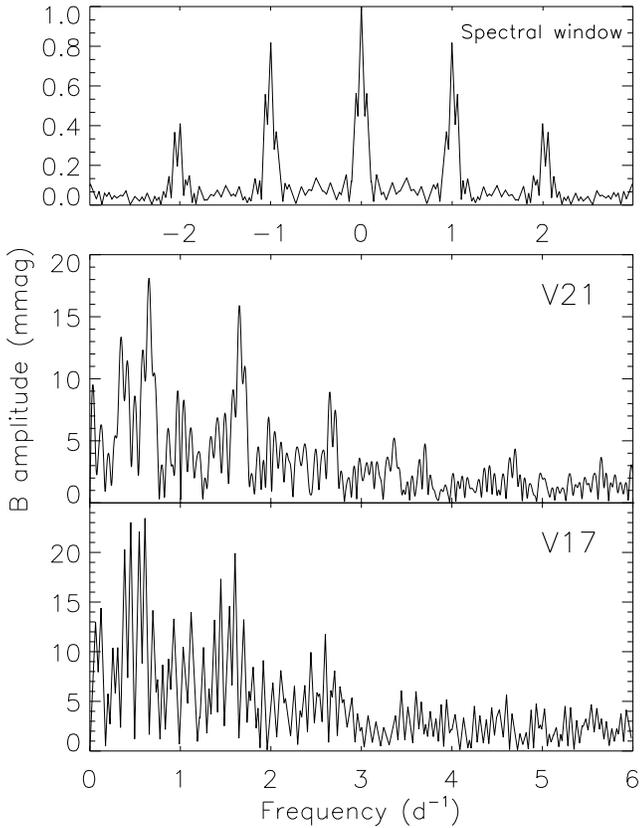}}
\caption{Spectral window function and examples of amplitude spectra for
two stars for which we detect long-period variability.
\label{fig.window}}
\end{figure}

In Fig.~\ref{fig.gdlc} we show light curves for 7 of the 15 $\gamma$~Doradus
candidates. They clearly display variability on time scales of days, and 
beating effects resulting from 
multiperiodicity is evidently present in several stars. 
The amplitude spectra of two of the stars are shown in Fig.~\ref{fig.window} 
(V17 and V21), along with the spectral window function shown at the 
same frequency scale. By comparing to the spectral window, which shows the 
signature of a single frequency in the amplitude spectrum, it is again 
evident that several frequencies are present, especially in V17. Indeed, 
we detect 6 and 3 frequencies for V17 and V21, respectively, 
in the range 0.1--2.5~\cd. 

\begin{figure} 
\resizebox{\hsize}{!}{\includegraphics{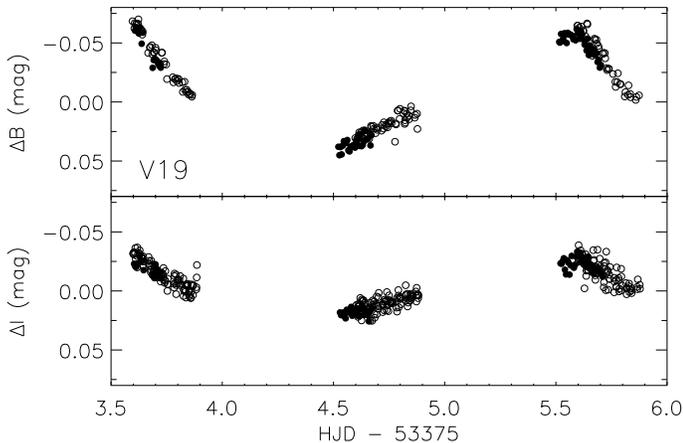}}
\caption{Three nights of merged data for the $\gamma$~Doradus star V19. 
Filled circles are data from La Palma, open circles
from La Silla. The amplitude difference between $B$ and $I$ is evident. 
\label{fig.gdmerging}}
\end{figure}

In Fig.~\ref{fig.gdmerging} we show an example of the merging of data 
from La Palma and La Silla for the variable V19. For 12 of the 15 stars
we have data from both sites and in all cases the overlapping data
agree well (if this was not the case, the star would be rejected from the 
list of variables). This excludes an instrumental origin of the long-period
variability, which must be intrinsic to the stars. For V22,V24 and V25 we
have data from La Silla only, but the variability is in all cases clearly
seen, and present in both the $B$- and $I$-filter data, leaving little room
for doubt about the intrinsic nature of the light variability. 

All detected frequencies are listed in Table~\ref{tab.gd}. We find 3--6 
frequencies in several of the stars, and only three stars appear mono-periodic
in our data. In general we find $I/B$ amplitude ratios of 
$0.4-0.6$ ($A_{I}/A_{B}$) while a wide range in phase differences are observed. 
The typical residual noise level in the amplitude spectra in the frequency 
range $0-6$~\cd~are $0.5-2.0$ mmag for these $V=15-16$ stars. 
These numbers explain why Kim et al. (2001) could not find $\gamma$ Doradus 
stars 
in their dataset. Indeed, their dataset consisted of 304 images, and they 
quote a detection limit of 10 mmag at V = 14. The corresponding number for our
dataset is 2 mmag, as residual noise levels in the amplitude spectra
of the V=14 stars in our sample are typically below 0.5 mmag.

\subsection{Are they $\gamma$~Doradus stars?}

To be considered a {\it bona fide} $\gamma$~Doradus star, 
a star should be located in the CMD inside
the $\gamma$~Doradus instability strip, be variable on time scales from 
$\sim0.4-3$ days, and all mechanisms for variability other than 
oscillations must be excluded. Such other mechanisms would typically be 
binarity (ellipsoidal variability) or rotational modulation (star spots).

All 15 stars are positioned inside the $\gamma$~Doradus instability strip
near the cluster main sequence turn-off. From isochrone fitting we find 
that the bluemost point on the cluster sequence corresponds to a mass of 
$1.4-1.5M_{\odot}$. This result is corroborated by a preliminary analysis 
of the binary V4, which places the primary at the top of the cluster main
sequence with a mass of $1.5\pm0.1M_{\odot}$ and the secondary further 
down the main sequence, below the instability strip, with a mass of 
$1.2\pm0.1M_{\odot}$ (see Fig.~\ref{fig.V4solve}). This indicates that 
the turn-off stars are of early $F$-type\footnote{Allen's Astrophysical 
Quantities, 4th edition, Arthur N. Cox, Ed.}, in good agreement with the 
$\gamma$~Doradus phenomenon occurring in stars of type $A7-F5$ (Kaye et al.
1999). The position of the 
15 variables near the cluster turn-off is thus fully in agreement with 
a $\gamma$~Doradus classification.

For the multiperiodic variables in our sample, mechanisms other than 
oscillations can readily be excluded as the cause of variability (see 
Handler 1999, Handler \& Shobbrook 2002 for a detailed discussion).
In short, ellipsoidal variables and rotationally modulated stars would be 
characterized by a single frequency, or by two frequencies with a harmonic 
relation. In the case of spots, two close frequencies may appear. For 
ellipsoidal variables, amplitude ratios between different passbands are 
expected to be unity, while $V/B$ amplitude ratios of 0.88-0.89 are expected
for starspot variables (Henry et al. 2005). For oscillations, this ratio 
is expected to be 0.67--0.80 (Garrido 2000) and typically 0.3--0.6 for $I/B$
(see Sect.~5.2), which is also what we find for most frequencies 
(Table~\ref{tab.gd}).
 
Among the 15 $\gamma$~Doradus stars, there are three mono-periodic stars,
but for both V13 and V24, further frequencies are likely present and only 
just escape detection in our data set. For V13 and V24, the amplitude ratios 
are also near 0.4, pointing towards oscillations and clearly rejecting 
ellipsoidal variability. The near detection of further frequencies, along 
with the low $I/B$ amplitude ratios suggest that also V13 and V24 are 
oscillating. For V12, the amplitude ratio is 0.7 which rejects ellipsoidal 
variability.
V13 (along with V12) are in the Mercator field and the data from March and 
April show that the variability is coherent over time scales of many months.
Star spots are not expected in $\gamma$~Doradus stars due to the early 
F-type classification (Kaye et al. 1999, Henry et al. 2005), except if they
are chemically peculiar. As we will show below, none of the three monoperiodic
stars appear chemically different from the rest of the group based on the 
Str{\"o}mgren photometry, but all 15 stars have low values of $\delta c_1$, 
which is an indicator of chemical 
peculiarity (Joshi et al. 2006). Thus, we cannot completely rule out that 
V12 could be a rotationally modulated Ap star, this can only be done 
spectroscopically. 
For V22, for which we do not have Str{\"o}mgren indices except for the position
in the CMD inside the instability strip, $\nu_{1}$ and $\nu_{2}$ are quite 
close to be harmonically related, 
but the $I/B$ amplitude ratios of 0.4 excludes ellipsoidal variability. 
The evidence therefore suggest oscillations as the cause of variability 
for 14 of the 15 stars, in agreement with a $\gamma$~Doradus classification.

\subsection{Comparison with the known $\gamma$~Doradus stars}

\begin{figure} 
\resizebox{\hsize}{!}{\includegraphics{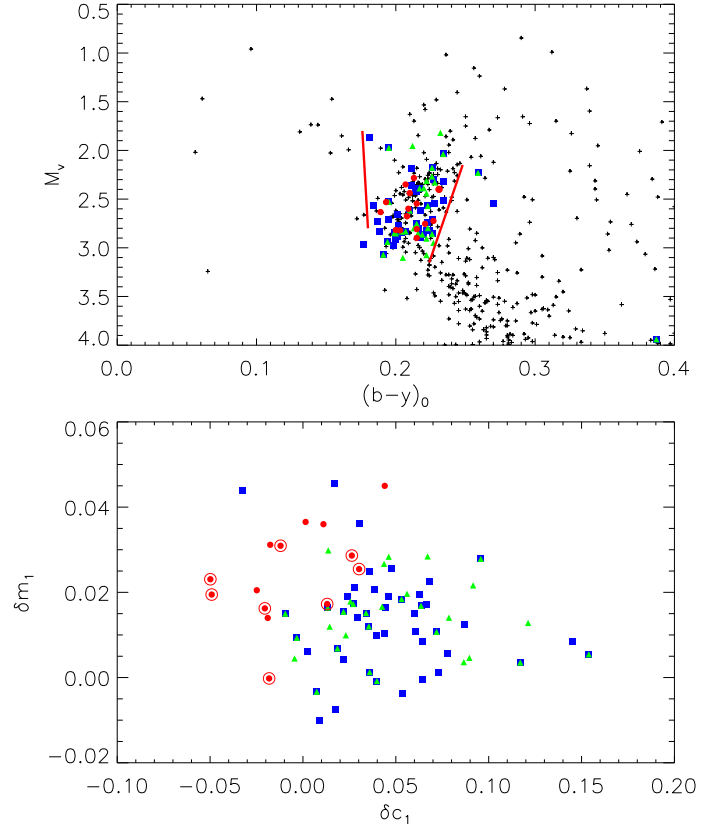}}
\caption{Comparison of stellar parameters between the NGC2506 $\gamma$~Doradus
stars and the known variables (Handler, private communication).
The {\it bona fide} $\gamma$~Doradus stars are shown as squares, the prime 
candidates as triangles, the NGC2506 stars as filled circles, and the 
instability strip is from Handler \& Shobbrook (2002). In the 
bottom panel, the multiperiodic (3 or more frequencies) NGC2506 stars have 
been marked with an open circle as well. 
\label{fig.compare}}
\end{figure}

We have compared the Str{\"o}mgren indices of the 15 NGC2506
variables to those of the {\it bona fide} and prime-candidate $\gamma$~Doradus
stars based on updated tables kindly provided by G. Handler. For the latter 
stars we extracted Str{\"o}mgren data from the catalogue of E.~H.~Olsen 
(private communucation)
and used the $b-y$ calibration of Crawford (1975) to derive $\delta c_1$
and $\delta m_1$. The result is shown in Fig.~\ref{fig.compare}. 

In the $M_V,(b-y)_0$ diagram,
the NGC2506 stars overlap completely with the known variables. In the 
($\delta c_1$,$\delta m_1$) diagram, the NGC2506 stars have a slightly 
more negative $\delta c_1$ and, in general, slightly higher $\delta m_1$. The 
higher $\delta m_1$ values suggest lower metallicity, in agreement with the 
metal content of the cluster, while the more negative $\delta c_1$ suggest
that the stars may be more chemically peculiar and therefore could be 
Ap stars (Joshi et al. 2006). The NGC2506 stars are, however, still located 
very close to the known $\gamma$~Doradus stars in this
($\delta c_1$,$\delta m_1$) diagram, where we have marked the variables with 
three or more frequencies with an open circle as well. The variability of 
those multiperiodic stars are certainly due to oscillations, and this subset
includes the two stars with the most negative values of $\delta c_1$.  
The 15 NGC2506 stars all have very similar indices; they form a homogeneous 
group in both
the CMD and the ($\delta c_1$,$\delta m_1$) diagram, and nothing suggests 
that any one star is chemically different from the others. We mention in
passing that this supports oscillations as the cause of variability in V12 
as well, as the star has indices similar to those of the 14 other 
$\gamma$~Doradus candidates.
The comparison with the known oscillators therefore strengthen the 
$\gamma$~Doradus classification of the NGC2506 stars. The minor shift of
the NGC2506 stars in $\delta c_1$,$\delta m_1$ may be due to a small
zero-point shift of our Str{\"o}mgren photometry compared to the known
variables, which are typically closer field-stars, or to an actual increase 
in the $\delta c_1$,$\delta m_1$ parameter space of the $\gamma$~Doradus stars 
in general.

The fact that all the NGC2506 $\gamma$~Doradus stars lie on or near the cluster 
main sequence inside the $\gamma$~Doradus instability strip is a strong 
indication that they are all cluster members. 
In fact, plots of various combinations of the Str{\"o}mgren indices given
in Table~\ref{tab.param}, all suggest that the $\gamma$~Doradus stars are 
cluster member turn-off stars, in full agreement with the CMD in 
Fig.~\ref{fig.cmd}. If some 
of the 15 stars were foreground or background objects, and thus were of a 
different spectral type, they would have Str{\"o}mgren indices separating 
them out from the large group of cluster turn-off stars.
It would also be difficult to explain the pulsational behaviour of such 
objects; background Cepheids or RR~Lyrae stars would have a characteristic 
light curve with a larger amplitude of variability. A reddened, background 
SPB variable could show the same type of variability as in the NGC2506 stars,
but would be unlikely to show Str{\"o}mgren indices identical 
to those of F-stars in a metal-poor cluster.

Furthermore, we have VLT/FLAMES spectra for V20, V21 and V24, of which the
first two are multiperiodic oscillators, in a data set of 120 stars in the 
cluster. We have derived preliminary radial velocities based on these spectra
(see Sect.~6), and the result is that at least V20 and V24 are cluster members 
as their mean radial velocity corresponds to the cluster mean velocity.
This provides additional, strong support for the classification 
of all 15 stars as cluster member $\gamma$~Doradus stars.
The velocities of V21 are in agreement with cluster membership but
long-term RV variations may be present. 

We therefore suggest that the NGC2506 variables V11--V25 except V12 should 
be included in the list of {\it bona fide} $\gamma$~Doradus stars. V12 should
be regarded as a prime candidate until the nature of the variability has been
established with certainty. This population of $\gamma$~Doradus stars 
makes NGC2506 extremely interesting for studying the $\gamma$~Doradus 
phenomenon, as these stars cover a large fraction of the instability strip 
and at the same time are located very near to the evolutionary point of 
turn-off from the main sequence.

We finally note that we did not find $\delta$~Scuti oscillations 
in any of the $\gamma$~Doradus stars, despite the fact that many of them
are inside the $\delta$~Scuti instability strip as well. This supports the
idea that $\gamma$~Doradus and $\delta$~Scuti oscillations are rarely present
in the same star (Handler \& Shobbrook 2002, Henry et al. 2005).
Both types of oscillation have, however, been found in HD8801 
(Henry \& Fekel 2005) and in BD+184914 (Rowe et al. 2007)  -- and possibly in 
V3 in this paper.
It is interesting to note that we do not find $\delta$~Scuti 
oscillations in {\it any} of the turn-off stars, despite many candidates inside 
the $\delta$~Scuti instability strip. In contrast, the open cluster NGC1817
also has turn-off inside (in the middle of) the $\delta$~Scuti instability 
strip, but that cluster,
which is half as old as NGC2506, is hosting a population of at least 12 
turn-off $\delta$~Scuti stars (Arentoft et al. 2005).
  
\begin{table}
\small
\caption{\normalsize{Variability data for the $\gamma$~Doradus stars. 
Frequencies are in \cd, amplitudes in mmag and phases in degrees.
The quoted errors on the frequencies, amplitude ratios and phase 
differences are standard deviations on the values determined from extensive 
simulations (see text).}
\label{tab.gd}}
\begin{tabular}{cccrrrr}
\hline
\noalign{\smallskip}
ID & $\nu_{i}$ & $\nu$ & A$_{B}$ & S/N$_{B}$ & A$_{I}$/A$_{B}$ & $\phi_{I}-\phi_{B}$ \\
\noalign{\smallskip} \hline
V11& $\nu_{1}$ & 1.165(1) & 7.93 & 6.78 & 0.41(4)  & -1(5)  \\
   & $\nu_{2}$ & 1.270(1) & 5.12 & 4.37 & 0.40(6)  &  6(8) \\
   & $\nu_{3}$ & 1.400(2) & 8.58 & 7.34 & 0.40(3)  &  3(5)  \\ 
V12& $\nu_{1}$ & 1.395(4) & 9.26 & 5.84 & 0.70(6)  &  30(6)  \\
V13& $\nu_{1}$ & 0.830(1) & 8.10 & 6.00 & 0.39(4)  &  13(6) \\
V14& $\nu_{1}$ & 0.758(4) & 2.130& 4.57 & 0.68(13) &  33(12) \\
   & $\nu_{2}$ & 0.870(1) &17.750&38.08 & 0.43(2)  &  -8(2)  \\
   & $\nu_{3}$ & 0.927(1) & 4.090& 8.77 & 0.31(8)  &  88(18) \\
V15& $\nu_{1}$ & 0.495(4) & 3.60 & 3.16 & 0.46(13) &  35(17) \\
   & $\nu_{2}$ & 0.614(1) &14.93 &13.10 & 0.48(2)  &   4(2)  \\
   & $\nu_{3}$ & 1.130(1) & 6.28 & 5.51 & 0.45(4)  & 131(5)  \\
V16& $\nu_{1}$ & 0.474(20) & 9.599& 5.02 & 0.29(3)  &  14(13) \\
   & $\nu_{2}$ & 0.795(1) &20.603&10.78 & 0.52(2)  & -13(2)  \\
   & $\nu_{3}$ & 0.918(3) &11.459& 6.00 & 0.49(4)  &   1(4)  \\
   & $\nu_{4}$ & 1.730(13) & 6.144& 3.21 & 0.73(11) &  -4(8)  \\
V17& $\nu_{1}$ & 0.136(1) &10.875& 7.64 & 0.68(4) &   21(3)  \\
   & $\nu_{2}$ & 0.606(1) &28.233&19.85 & 0.46(2) &    4(2)  \\
   & $\nu_{3}$ & 0.645(7) &15.658&11.01 & 0.51(3) &  -11(3)  \\
   & $\nu_{4}$ & 1.268(1) & 7.362& 5.17 & 0.55(5) &   -3(5)  \\
   & $\nu_{5}$ & 1.537(2) & 7.213& 5.07 & 0.35(11) &  28(15) \\
   & $\nu_{6}$ & 2.439(1) & 5.711& 4.01 & 0.65(8) &   -2(7) \\
V18& $\nu_{1}$ & 0.206(1) & 8.939& 9.51 & 0.80(4) &  -15(3) \\
   & $\nu_{2}$ & 0.674(2) & 3.767& 4.01 & 0.96(11)&  -14(6)\\
V19& $\nu_{1}$ & 0.156(1) & 7.190& 6.13 & 0.56(4) &    1(4) \\
   & $\nu_{2}$ & 0.442(2) &15.774&13.44 & 0.51(2) &    1(2) \\
   & $\nu_{3}$ & 0.583(1) &20.767&17.70 & 0.39(2) &    0(3) \\
   & $\nu_{4}$ & 1.344(3) & 6.928& 5.90 & 0.51(5) &  -27(5) \\
   & $\nu_{5}$ & 1.593(5) & 6.902& 5.88 & 0.51(6) &    5(6) \\
V20& $\nu_{1}$ & 0.842(25) & 7.610&11.23 & 0.43(5) &  -28(7) \\
   & $\nu_{2}$ & 0.883(1) & 9.963&14.70 & 0.66(5) &   -2(3) \\
   & $\nu_{3}$ & 1.701(16) & 3.011& 4.44 & 0.47(8) &  -28(11)\\
V21& $\nu_{1}$ & 0.503(1) & 5.069& 3.50 & 0.53(9) &    4(15)\\
   & $\nu_{3}$ & 0.648(1) &17.793&12.30 & 0.40(2) &  -15(2) \\
   & $\nu_{3}$ & 1.740(1) & 5.434& 3.76 & 0.60(5) &   29(6) \\
V22& $\nu_{1}$ & 0.700(1) &14.214& 6.18 & 0.44(3) &   -8(4) \\
   & $\nu_{2}$ & 1.378(1) &12.684& 5.52 & 0.42(3) &    2(4) \\
V23& $\nu_{1}$ & 1.168(3) & 9.616& 6.35 & 0.47(4) &    1(5) \\
   & $\nu_{2}$ & 1.260(3) &14.415& 9.52 & 0.52(3) &   12(3) \\
V24& $\nu_{1}$ & 0.916(1) &14.458&12.25 & 0.45(3) &   16(5) \\
V25& $\nu_{1}$ & 1.178(1) &14.747&10.84 & 0.60(3) &   12(2) \\
   & $\nu_{2}$ & 1.987(2) & 7.055& 5.19 & 0.50(5) &  -18(8)\\
\hline
\noalign{\smallskip}
\end{tabular}
\end{table}

\subsection{Comparison with theoretical predictions}

As mentioned in the introduction, $\gamma$ Doradus stars are rich stellar 
laboratories to study astrophysical phenomena. This, however, makes them
complex and challenging targets for which we need as many additional 
observational constraints as possible. 
Such constraints will be obtained when we determine with high precision the
masses and radii of the two eclipsing binaries in NGC2506. The
high-resolution spectroscopic datasets involved, obtained with 
the VLT at Paranal Observatory, are currently being reduced and analysed.
The datasets will pinpoint the cluster isochrone with an unprecedented 
precision for NGC2506, which will lead to masses, 
radii and the age of all stars along the cluster main sequence, including the 
$\gamma$ Doradus stars. In addition, a detailed abundance analysis of
cluster stars based on VLT/FLAMES spectra will shed light on the metallicity 
of the cluster stars. 
These results, which will be published in a separate paper, 
will provide the needed, additional constraints and we therefore 
postpone a detailed seismic analysis of the $\gamma$ Doradus candidates 
until they are available. 

\begin{figure}
\resizebox{\hsize}{!}{\includegraphics{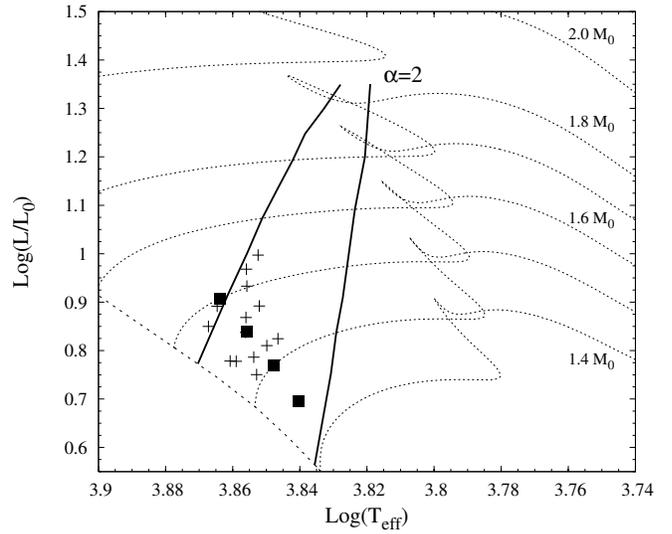}}
\caption{An HR-diagram showing the 4 generic $\gamma$ Doradus models as 
squares together with the positions of the $\gamma$ Doradus candidates as crosses. 
Also shown are the ZAMS, evolutionary tracks and the blue and red border of 
the $\gamma$ Doradus instability strip for $\ell=1$ modes and for the mixing 
length parameter $\alpha=2$. \label{theomodels}}
\end{figure}

We did, however, perform a preliminary theoretical modeling to verify 
whether there is at least a crude correspondence between the observed 
amplitude ratios and phase differences and the theoretical ones. Rather 
than modeling each target individually and being subjected to the possible 
large, systematic uncertainties on their positions in the HR-diagram, we 
choose to compute 4 generic stellar models using the stellar evolution 
code CL\'eS (R. Scuflaire, University of Li\`ege, private communication). 
Each of these models was computed for the same 
cluster age as specified in Section \ref{targetcluster} but with different 
effective temperatures going from the blue to the red border of the 
$\gamma$ Doradus instability strip. These generic models, together with the 
positions of the $\gamma$ Doradus candidates are shown in the HR-diagram in 
Fig.~\ref{theomodels}.

Consequently, we computed the $\ell = 1, 2$ and $3$ eigenfrequency spectra 
for each of the generic models using the non-adiabatic oscillation code MAD. 
This code uses time-dependent convection (TDC) theory as described by 
Grigahc\`ene et al.~(2005) and Dupret et al.~(2005a), and was successfully 
applied to five $\gamma$ Doradus field stars by Dupret et al.~(2005b). 
From these spectra, and for each degree $\ell$, we selected the 
modes with eigenfrequencies closest to the observed frequencies and computed 
the non-adiabatic relative amplitude and phase 
of the $T_{\rm eff}$ and gravity variation. These quantities were then 
used to compute the photometric amplitude ratio $I/B$ and phase difference 
$I-B$. The theoretical model used is the same as the one used in Dupret 
et al.~(2005b), and incorporates Kurucz (1993) atmosphere 
models and the 4-coefficient non-linear limbdarkening law of Claret (2000). 
Through interpolation in metallicity grids, the results were computed for the 
cluster metallicity mentioned in Section \ref{targetcluster} ([Fe/H] = $-0.20$).

Three important results come out from our comparison between theoretical 
predictions and observations. First, the generic models are unable to 
explain the observed oscillation frequencies below 0.25 d$^{-1}$.
In particular, $\nu_1 = 0.136$ d$^{-1}$ of V17,  $\nu_1 = 0.206$ d$^{-1}$ of
V18, and $\nu_1 = 0.156$ d$^{-1}$ of V19 are predicted to be stable 
(i.e. not excited).
The most important parameter determining the (in)stability of eigenmodes 
in $\gamma$ Doradus stars is the mixing-length parameter $\alpha$. We used the
value $\alpha=2.0$ because a) testing revealed that this value leads to the
largest number of unstable modes, and b) Dupret et al.~(2005b) successfully
used this value to explain best the observations of five field $\gamma$ Doradus 
stars. But in the case of our cluster targets this value seems not able to 
explain the complete observed frequency spectrum. We remark, however, that the 
observed frequency $\nu_3 = 0.343$ d$^{-1}$ of the well known $\gamma$ Doradus star 
9 Aur is also predicted to be stable by Dupret et al.~(2005b).

\begin{figure}
\resizebox{\hsize}{!}{\includegraphics{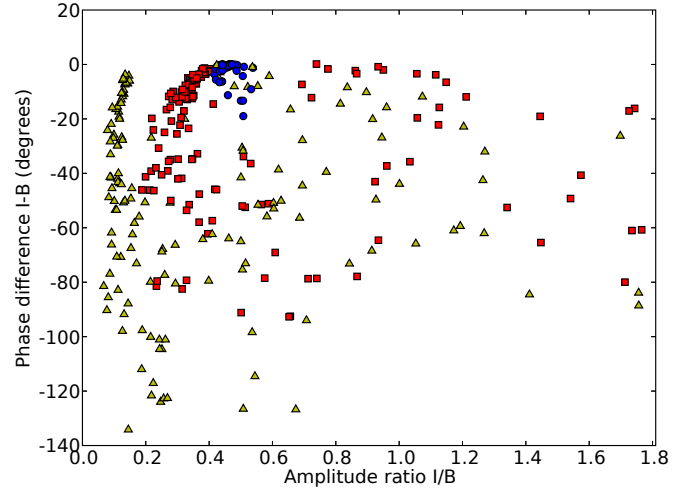}}
\caption{The theoretical photometric amplitude ratios $I/B$ versus phase 
difference $I-B$. It concerns the modes of the 4 generic models shown in 
Fig.~\ref{theomodels} for which the eigenfrequency is close to one of the 
observed frequencies in Table \ref{tab.gd}. The blue circles denote 
$\ell=1$ modes, the red squares $\ell = 2$ modes and the yellow triangles 
$\ell = 3$ modes. The results were computed for the cluster metallicity 
stated in Section \ref{targetcluster}, and for the mixing length parameter 
$\alpha=2$. \label{theo_ratiophase}}
\end{figure}

The second result concerns the photometric phase differences $I-B$. 
Fig.~\ref{theo_ratiophase} shows the theoretically predicted photometric 
amplitude ratios $I/B$ and corresponding phase difference $I-B$ 
for all modes of all targets, assuming degrees $\ell = 1, 2$ and $3$. 
Comparing this figure with the corresponding observational 
Fig.~\ref{ratiodiff} reveals a striking difference: for many of the 
oscillation modes we observe a \emph{positive} phase difference $I-B$ which 
cannot be explained by the models. The phase difference of these modes is 
usually positive well within the error bars, and this is even so for several 
of the highest-amplitude modes for which no aliasing-problems was detected 
(red triangles). There is little to compare to in the literature but we 
mention the relevant information in the forthcoming paper of 
Cuypers et al.~(2007, in preparation). In this paper a sample of 
19 $\gamma$ Doradus field stars is presented that was
photometrically monitored in the Geneva passbands. The authors always 
find a phase difference between the $U$ and $G$ filter that is very close to 
zero (within the standard errors). This turns out to be in agreement
with the theory. Indeed, we performed similar simulations for the Geneva
$U$ and $G$ filters, and we found that the phase difference $U-G$ behaves 
somewhat
different than $I-B$. First, it can theoretically be both positive and negative.
Secondly, regardless the degree, the phase difference $U-G$ is almost always
close to zero. The results of Cuypers et al.~are therefore not surprising. It
does seem to show, however, that the phase difference $I-B$ is more suitable
for mode identification as it has more discriminating power.

A third remark worth mentioning is the scatter of the observed modes in 
Fig.~\ref{ratiodiff}. Intuitively we would expect more $\ell=1$ modes to 
be present among our sample as a selection effect because these modes 
suffer less surface cancellation. This effect seems to be present
in the sample of Dupret et al.~(2005b) who identified 8 out of 12 
studied $\gamma$ Doradus modes unambiguously as dipole modes. Note, however, 
that the $\ell=1$ modes in Fig.~\ref{theo_ratiophase} (blue circles) occupy 
a remarkably small region given the fact that the 4 generic models from 
which they were computed are spread over a significant part of the $\gamma$ 
Doradus instability strip. The rather large scatter in the observational 
diagram in Fig.~\ref{ratiodiff} therefore shows that the majority of our 
sample of $\gamma$ Doradus modes cannot easily be identified as dipole modes.

Given the evidence, we should conclude that either the observational 
uncertainties on the photometric amplitude ratios and phase differences 
are significantly larger than we estimated, or the theoretical model is 
inadequate. We verified that changing the metallicity (e.g.~to [Fe/H] = $0.0$ 
or [Fe/H] = $-0.3$) does not change qualitatively our results. 
In our forthcoming theoretical study we will explore a larger parameter 
space where we will examine in more detail the influence of the overshoot 
parameter, and the rotational velocity. The latter in particular may be 
important for $\gamma$
Doradus stars as for these stars the rotation over oscillation frequency ratio 
can easily be comparable to or larger than 1.

\section{Eclipsing Binaries}
\label{EclipsingBinaries}

\begin{figure} 
\resizebox{\hsize}{!}{\includegraphics{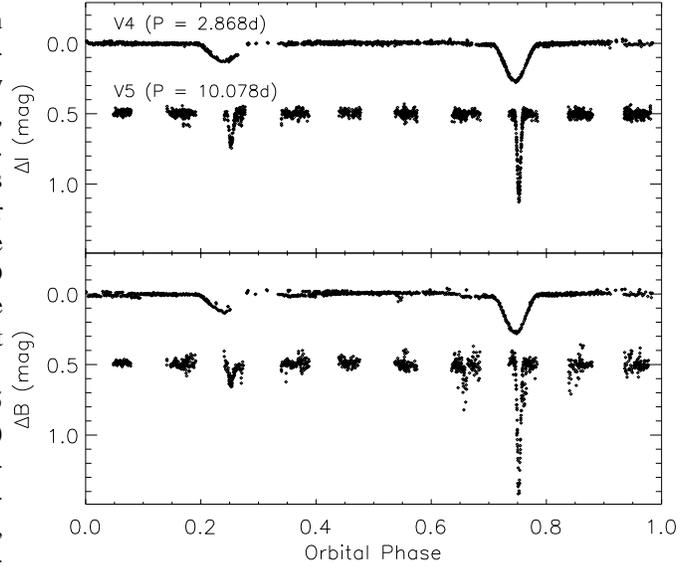}}
\caption{Phased light curves for the detached binaries V4 and V5.
\label{fig.V4V5phased}}
\end{figure}

\begin{figure} 
\resizebox{\hsize}{!}{\includegraphics{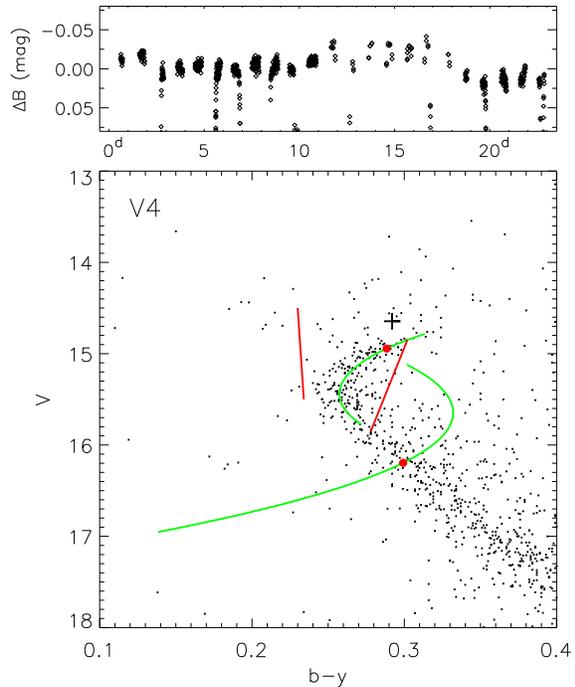}}
\caption{24-day light curve (top) and photometric solutions to the
V4 binary system. The solid red lines are the $\gamma$~Dor instability strip.
The cross denotes the position of V4 in the CMD and the green curves are 
possible photometric solutions of the two components. The red dots mark a 
solution that puts both components on the cluster sequence.
\label{fig.V4solve}}
\end{figure}

As mentioned above, NGC2506 hosts the two detached, eclipsing binaries V4 and 
V5. We have used the present photometric data set to determine, for 
the first time, the periods of these two binaries: 2.868 and 
10.078 days for V4 and V5, respectively. The phased light curves are shown in 
Fig.~\ref{fig.V4V5phased}, including some nights of poor quality. 
The $B$ light-curve of V4 display minor offsets between data taken at the 
same phase, but on different nights. The same is the case for $I$, although
to a lesser extent. 
Fig.~\ref{fig.V4solve} illustrates this in more detail. 
If we assume that the binary is a cluster
member, as indicated by the spectroscopic data to be discussed below, the
primary component should lie on the cluster sequence near
turn-off, i.e., somewhere along the uppermost of the green curves in 
Fig.~\ref{fig.V4solve}, lower panel. If we also require
the secondary to be on the cluster sequence, and that the combined 
light of the two main sequence stars should be equal to the observed, total 
light, we obtain a photometric solution that puts the primary inside the 
$\gamma$~Doradus instability strip and the secondary somewhat below it, as 
indicated by the red dots in the diagram. In the upper panel of 
Fig.~\ref{fig.V4solve} we show the light curve in more detail.
Additional variability is present on a time scale of several days which is,
however, longer than the typical $\gamma$~Doradus type variability. Further
discussion of this object is therefore postponed until all data, including
the spectroscopy, are available.

\begin{figure} 
\resizebox{\hsize}{!}{\includegraphics{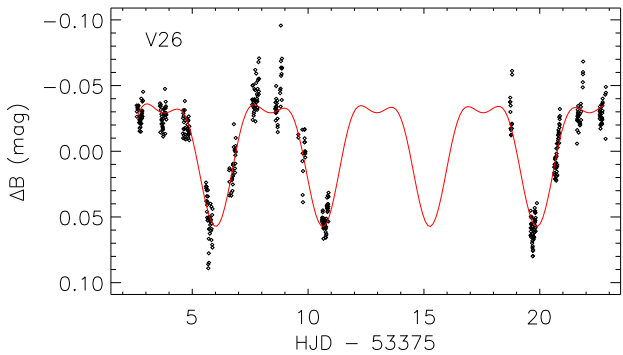}}
\caption{The new probable binary V26. The red curve shows a 2-frequency 
fit with a 1:2 relationship between them.
\label{fig.V26}}
\end{figure}

\begin{figure} 
\resizebox{\hsize}{!}{\includegraphics{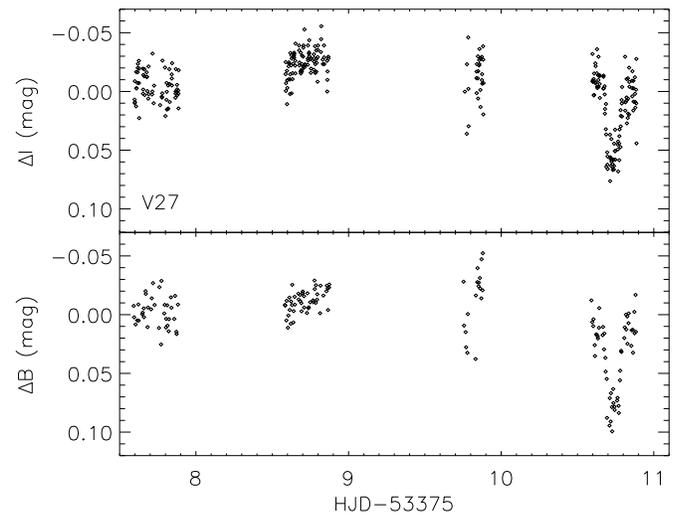}}
\caption{The new probable binary V27; four nights of data are shown, with
a possible eclipse taking place during the last night. 
\label{fig.V27}}
\end{figure}

\begin{figure} 
\resizebox{\hsize}{!}{\includegraphics{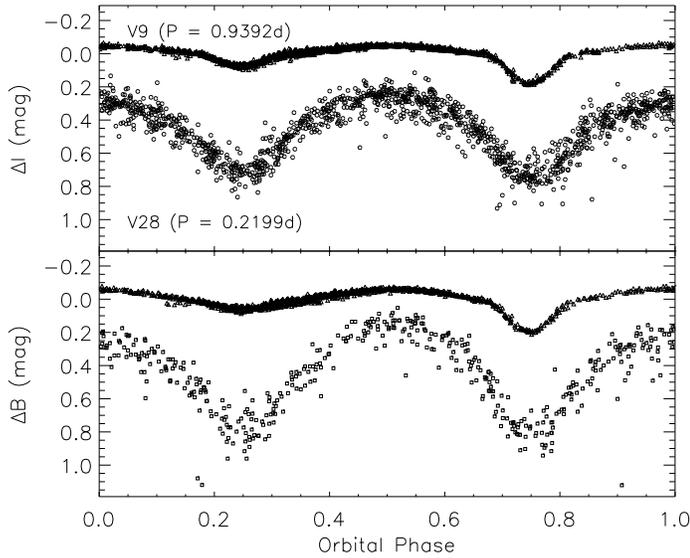}}
\caption{Phased light curves for the new binaries V9 and V28. The latter 
object is not a cluster member based on the position in the CMD.
\label{fig.V9V28phased}}
\end{figure}

\begin{figure*}[h] 
\resizebox{\hsize}{!}{\includegraphics{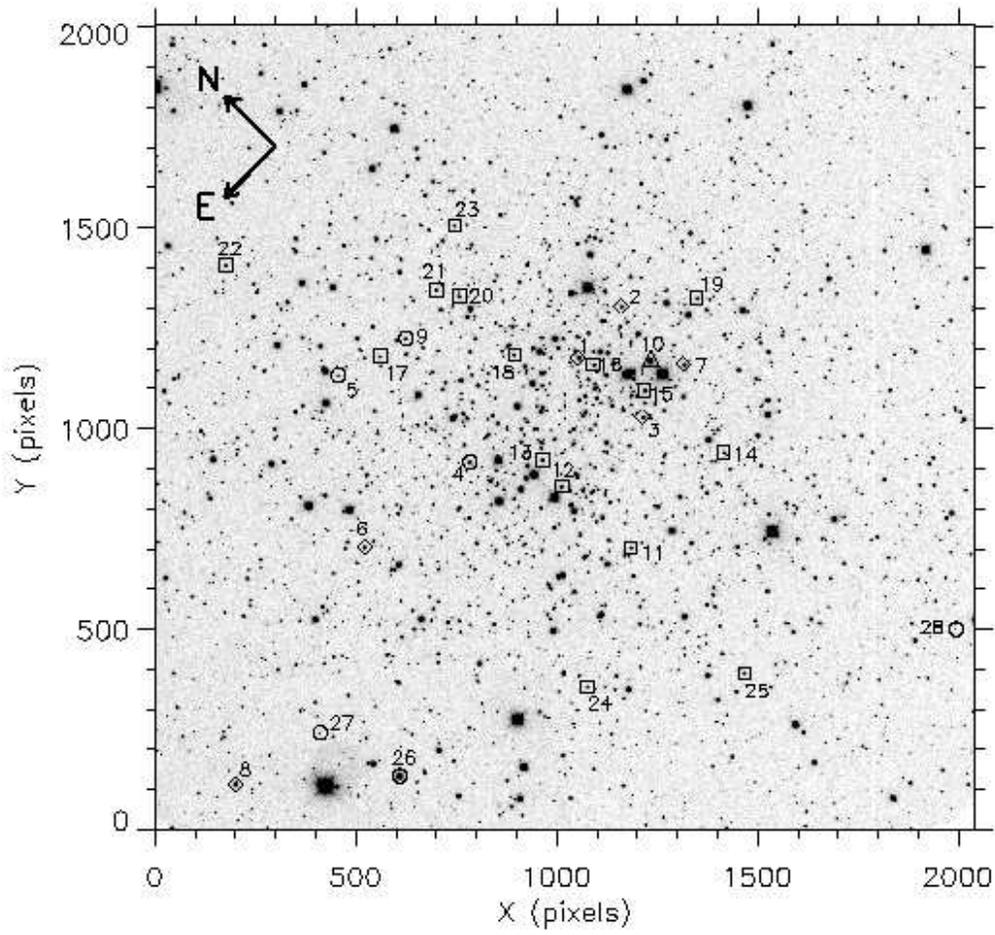}}
\caption{DFOSC image serving as a finding chart for NGC2506, with the 
variables indicated. The image orientation is indicated by arrows. 
\label{fig.finding}}
\end{figure*}

From the photometric data, we also detect four new eclipsing binaries 
(V9 and V26-V28). Light curves are shown in Figs.~\ref{fig.V26}, ~\ref{fig.V27}
and~\ref{fig.V9V28phased} and we refer to Table~\ref{tab.param} for photometric
indices. V26 is a giant star, near the limit of saturation
in our data, especially in I. Fig.~\ref{fig.V26} plots the evidence of 
binarity. V27 (Fig.~\ref{fig.V27}) is possibly a new, detached eclipsing 
binary positioned in the CMD on the cluster main sequence. A single, probable
eclipse was observed in both $B$ and $I$ during one night. 
Finally, we detect the two binaries V9 and V28 (Fig.~\ref{fig.V9V28phased}). 
V9 is located in the CMD inside the $\gamma$~Doradus instability strip,
near the cluster sequence, and may be an Algol- or $\beta$~Lyrae type binary. 
V28 is faint, red and, from the position in the CMD far from the cluster 
main sequence, not a cluster member. The short period (0.2199 days), and the
light-curve shape, suggest a W UMa classification.

The spectroscopic data-set mentioned above consists of 15 
epochs of VLT UVES spectra for 
5 stars (V4 and four giants) and Giraffe/FLAMES spectra of 120 stars 
(including V5) observed in service mode.  
The aims of these data are to analyse the binary stars for 
deriving precise, absolute parameters, as well as to use the Giraffe 
spectra to determine metallicity/abundances, ages and cluster membership. 
The latter allows a cleaning of the bright part of the CMD for non-members, 
which in turn allows much stronger constraints to be put on the isochrones.
The analysis of the NGC2506 spectra is still ongoing and will be the subject 
of a later paper, but we mention that a preliminary analysis of the UVES 
spectra shows that V4 is a double-lined binary and a cluster member.

We have also done a preliminary reduction of the Giraffe spectra to derive 
approximate radial velocities for the 120 observed stars, among these V5 and 
V9. Both stars show clear velocity variations centered on
the cluster mean value of approximately $+84\rm{\,km/s}$.
The two $\gamma$~Doradus stars V20 and V24, for which we
have Giraffe spectra, have constant velocities over the 15 epochs, with 
values near the cluster mean. They are thus cluster members, as are V4, 
V5 and V9.
For V21, long-term RV variations are possibly present, with values 
between $+70$ and $+90\rm{\,km/s}$, but further analysis is needed before any 
firm conclusions can be drawn about the membership status and possible 
dublicity of this object.

\begin{table*}
\scriptsize
\caption{Stellar parameters for the variables in NGC2506. Type 
names refer to {\it oscillating blue straggler} for BS, {\it detached eclipsing
binary} for DEB, {\it contact binary} 
\newline
for CB, {\it $\gamma$~Doradus} for GD,
and {\it eclipsing binary} (unspecified) for EB. V10 has not been classified, 
and Str{\"o}mgren photometry is not available for V8, V22 and V28.
\label{tab.param}}
\begin{tabular}{cccccccccccccccl}
\hline
\noalign{\smallskip}
ID & $\alpha_{2000}$ & $\delta_{2000}$ & $V$ & $b-y$ & $v-y$ & $m_{1}$ & $c_{1}$ & $(b-y)_{0}$ & $c_{0}$ & 
$T_{\rm eff}$ & $M_{V}$ & $M_{bol} $ & $BC$ & log$L$ & Type  \\
\noalign{\smallskip} \hline
V1 & 7 59 58.1&-10 45 54 & 13.660 &  0.150 &  0.468 &  0.168 &  0.994 &
0.096 & 0.983             & 8330   &  0.965 &  0.903 & -0.062 &  1.537 & BS \\ 
V2 & 7 59 53.6& -10 45 48 & 14.510 &  0.185 &  0.548 &  0.178 &  0.819 & 
0.131 & 0.808             & 7950   &  1.815 &  1.783 & -0.031 &  1.185 & BS \\ 
V3 & 7 59 57.9& -10 47 20 & 14.727 &  0.207 &  0.561 &  0.147 &  0.820 & 
0.153 & 0.809             & 7740   &  2.032 &  2.009 & -0.022 &  1.095 & BS \\
V4 & 8 00 08.2& -10 45 50 & 14.645 &  0.292 &  0.706 &  0.122 &  0.586 & 
0.238 & 0.575             & 6920   &  1.950 &  1.900 & -0.050 &  1.138 & DEB \\ 
V5 & 8 00 10.3& -10 43 17 & 17.430 &  0.456 &  1.133 &  0.221 &  0.199 &  
0.402 & 0.188             &        &        &        &        &        & DEB \\ 
V6 & 8 00 17.2& -10 45 35 & 14.677 &  0.243 &  0.701 &  0.215 &  0.651 & 
0.189 & 0.640             & 7360   &  1.982 &  1.965 & -0.017 &  1.113 & BS \\ 
V7 & 7 59 53.4& -10 47 12 & 14.694 &  0.220 &  0.599 &  0.159 &  0.752 & 
0.166 & 0.741             & 7600   &  1.999 &  1.983 & -0.015 &  1.105 & BS \\ 
V8 & 8 00 34.5& -10 46 50 & 13.80  &  0.22  &        &        &        &
      &                   &        &        &        &        &        & BS \\ 
V9 & 8 00 05.4& -10 43 39 & 15.289 &  0.268 &  0.666 &  0.130 &  0.681 & 
0.214 & 0.670             & 7160   &  2.594 &  2.563 & -0.030 &  0.873 & EB \\ 
V10& 7 59 54.9& -10 46 46 & 13.079 &  0.522 &  1.287 &  0.243 &  0.513 & 
0.468 & 0.502             &        &        &        &        &        &  ?  \\
V11& 8 00 04.6& -10 48 43 & 15.454 &  0.275 &  0.681 &  0.131 &  0.621 & 
0.221 & 0.610             & 7080   &  2.759 &  2.720 & -0.038 &  0.810 & GD \\ 
V12& 8 00 05.0& -10 47 11 & 15.508 &  0.269 &  0.654 &  0.116 &  0.654 & 
0.215 & 0.643             & 7140   &  2.813 &  2.780 & -0.032 &  0.786 & GD \\ 
V13& 8 00 04.7& -10 46 39 & 15.599 &  0.269 &  0.663 &  0.125 &  0.621 & 
0.215 & 0.610             & 7130   &  2.904 &  2.870 & -0.034 &  0.750 & GD \\ 
V14& 7 59 55.8& -10 48 41 & 15.376 &  0.262 &  0.687 &  0.163 &  0.610 & 
0.208 & 0.599             & 7180   &  2.681 &  2.652 & -0.029 &  0.838 & GD \\ 
V15& 7 59 56.5& -10 47 03 & 14.983 &  0.267 &  0.676 &  0.142 &  0.566 & 
0.213 & 0.555             & 7120   &  2.288 &  2.253 & -0.034 &  0.997 & GD \\ 
V16& 7 59 57.7& -10 46 09 & 15.333 &  0.243 &  0.624 &  0.138 &  0.668 & 
0.189 & 0.657             & 7370   &  2.638 &  2.621 & -0.017 &  0.850 & GD \\ 
V17& 8 00 07.4& -10 43 33 & 15.521 &  0.257 &  0.662 &  0.148 &  0.621 & 
0.203 & 0.610             & 7230   &  2.826 &  2.802 & -0.024 &  0.778 & GD \\ 
V18& 8 00 01.0& -10 45 06 & 15.517 &  0.254 &  0.642 &  0.134 &  0.632 & 
0.200 & 0.621             & 7260   &  2.822 &  2.800 & -0.021 &  0.778 & GD \\ 
V19& 7 59 49.7& -10 46 35 & 15.050 &  0.261 &  0.662 &  0.140 &  0.581 & 
0.207 & 0.570             & 7180   &  2.355 &  2.326 & -0.029 &  0.968 & GD \\ 
V20& 8 00 00.8& -10 43 47 & 15.138 &  0.264 &  0.673 &  0.145 &  0.636 & 
0.210 & 0.625             & 7170   &  2.443 &  2.414 & -0.029 &  0.933 & GD \\ 
V21& 8 00 01.6& -10 43 27 & 15.424 &  0.281 &  0.695 &  0.133 &  0.610 & 
0.227 & 0.599             & 7020   &  2.729 &  2.685 & -0.044 &  0.824 & GD \\ 
V22& 8 00 10.3& -10 40 42 & 15.10  &  0.29  &        &        &        &
      &                   &        &        &        &        &        & GD \\ 
V23& 7 59 57.7& -10 42 55 & 15.246 &  0.269 &  0.685 &  0.147 &  0.591 & 
0.215 & 0.580             & 7110   &  2.551 &  2.516 & -0.035 &  0.892 & GD \\ 
V24& 8 00 13.3& -10 49 49 & 15.299 &  0.263 &  0.652 &  0.126 &  0.627 & 
0.209 & 0.616             & 7180   &  2.604 &  2.575 & -0.029 &  0.868 & GD \\ 
V25& 8 00 05.3& -10 51 29 & 15.231 &  0.247 &  0.641 &  0.147 &  0.644 &
0.193 & 0.633             & 7320   &  2.536 &  2.517 & -0.019 &  0.892 & GD \\ 
V26& 8 00 26.5& -10 48 39 & 13.058 &  0.632 &  1.631 &  0.367 &  0.314 &
0.578 & 0.303             &        &        &        &        &        & EB \\ 
V27& 8 00 28.2& -10 47 13 & 17.175 &  0.368 &  0.912 &  0.176 &  0.219 &
0.314 & 0.208             &        &        &        &        &        & EB \\ 
V28& 7 59 53.2& -10 53 26 & 18.60  &  1.50  &        &        &        &
      &                   &        &        &        &        &        & CB \\ 
\hline
\noalign{\smallskip}
\end{tabular}
\end{table*}

\section{Summary and conclusions}

Our two-colour, dual-site photometric campaign from La Palma and La Silla
was a success as we have quadrupled the number of known variables 
in the open cluster NGC2506. Based on our results we conclude that the
cluster contains (at least) 6 oscillating BS stars, 15 $\gamma$ Doradus stars, 
and 6 eclipsing binaries.

Prior to this work, only three short-period variables were known in 
NGC2506, and only one of them was classified as a oscillating BS star. 
We now know 6 oscillating BS stars, of which 5 are multi-periodic. 
One BS star (V3) shows both short- and long-period variability. 
To our knowledge, this would be the first BS star for which both p-mode 
and g-mode oscillations are discovered. Given the fact that we detected 
11 oscillations frequencies in this star, makes this blue straggler
a very promising
target for follow-up research. We therefore reached our goal of data gathering 
and selection of interesting targets suitable for future investigation.
We finally mention that a preliminary mode
identification seems to indicate that in four of the BS stars the fundamental
radial mode is excited, and for one star (V6) there is evidence of 
the fundamental and first overtone frequencies being excited.  

We report the discovery of 15 $\gamma$~Doradus stars of which most are 
multiperiodic, and at least 14 of them are {\it bona fide} $\gamma$~Doradus 
stars, while one is 
a prime $\gamma$~Doradus candidate. The $(b-y)$ colours, 
$M_V$ and the $\delta c_1$,$\delta m_1$ indices agree well with the 
corresponding values for the known $\gamma$~Doradus variables. 

We have compared frequencies, amplitude ratios and phase differences of the 
$\gamma$~Doradus stars to state-of-the-art $\gamma$~Doradus models. The 
results of this comparison are that 1) the lowest, detected frequencies
(below 0.25~\cd) are predicted to be stable in the models, 2) we observe
positive $I-B$ phase differences in disagreement with the models, and 
3) the majority of the observed frequencies appear not to be dipole modes, 
contrary with what is expected, as cancellation effects are stronger for modes 
of higher degree. These results clearly ask for further examination. We plan
a forthcoming theoretical study and additional spectroscopic data to 
investigate the effect of stellar rotation on the observables.

We have for the first time determined periods for the two detached, eclipsing 
binaries V4 and V5, and preliminary spectroscopic results confirms cluster 
membership for both. 

All these results combined strongly confirm and increase our interest in 
NGC2506 as an excellent target cluster for studying BS and $\gamma$ Doradus stars.

\acknowledgements
The Danish Natural Science Research Council, the Fund for Scientific
Research, Flanders, and the Instrument Center for Danish Astrophysics (IDA) 
are acknowledged for financial support. 
JDR and MR are postdoctoral fellows of the Fund for Scientific Research, 
Flanders, and FG acknowledges financial support from IDA.
This research has made use of the SIMBAD database, operated at CDS,
Strasbourg, France.

\end{document}